\documentclass[iop]{emulateapj}
\newcommand{\myemail}{lychiang@asiaa.sinica.edu.tw}
\newcommand{\bi}[1]{\mbox{\boldmath $#1$}}

\usepackage{float}
\usepackage{amsmath}
\usepackage{graphicx}
\usepackage{epsfig}
\usepackage{subfigure}
\usepackage{multirow}
\makeatletter

\def\alm{a_{\l m}}

\def\glesp{G{\sc lesp }}
\def\healpix{H{\sc ealpix }}

\def\planck{{\rm Planck }}
\def\wmap{{\rm WMAP }}

\def\etal{{et al.}}
\def\l{{\ell}}

\def\ylm{Y_{\l m}}

\def\sumlm{\sum_{\l m}}

\def\xk{{X_k}}

\slugcomment{Submitted to Astrophysical Journal}

\shorttitle{Excessive shift of the CMB acoustic peaks of the Cold Spot area 
}
\shortauthors{Lung-Yih Chiang}

\begin{document}
\title{Excessive shift of the CMB acoustic peaks of the Cold Spot area}
\author{Lung-Yih Chiang}
\affil{\asiaa}

\email{\myemail}

\newcommand{\asiaa}{{Institute of Astronomy and Astrophysics, Academia Sinica, 11F of Astro-Maths Building, AS/NTU. No.1, Sec.4, Roosevelt Rd, Taipei 10617, Taiwan}}

\begin{abstract}

Measurement of the acoustic peaks of the cosmic microwave background (CMB) temperature anisotropies has been instrumental in deciding the geometry and content of the universe. Acoustic peak positions vary in different parts of the sky owing to statistical fluctuation. We present the statistics of the peak positions of small patches from ESA Planck data. We found that the peak positions have significantly high variance compared to the 100 CMB simulations with best-fit $\Lambda$CDM model with lensing and Doppler boosting effects included. Examining individual patches, we found the one containing the mysterious "Cold Spot", an area near the Eridanus constellation where the temperature is significantly lower than Gaussian theory predicts, displays large synchronous shift of peak positions towards smaller multipole numbers with significance lower than $1.11\times 10^{-4}$. The combination of large synchronous shifts in acoustic peaks and lower than usual temperature at the Cold Spot area results in a 4.73$\sigma$ detection (significance $p\simeq 1.11\times 10^{-6}$) against the $\Lambda$CDM model. And it was already reported in Finelli et al. (2016) that in the WISE-2MASS galaxy catalog at $z<0.3$ the Cold Spot region is surrounded by surprisingly large underdense regions around 15$^\circ$ in radius, which are found to be in the same square patch. Thus we propose there is some extra localized unknown energy to stretch out the space in the transverse direction around the Cold Spot area to simultaneously account for the Cold Spot, the excessive shift of the acoustic peaks, and the large underdense regions.

\end{abstract}

\section{Introduction}
%The measurement of the acoustic peak positions of the cosmic microwave background temperature anisotropies has been instrumental in deciding the geometry and content of the universe, e.g. the first acoustic peak at $\l\simeq$220 indicating a flat universe. It was the Boomerang balloon experiment in 2000 that firstly shows the first acoustic peak position (in a partial sky) is measured at $\l=197\pm 6$ \citep{boomerang}, and later MAXIMA-1 at $\l\simeq220$ \citep{maxima}, compared to the latest \planck full-sky result fitted to be $\l\simeq$222 \citep{planckps}. The difference is due to the fact that the acoustic peak positions vary in different parts of the sky. 

The measurement of the acoustic peak positions of the cosmic microwave background (CMB) temperature anisotropies has been instrumental in deciding the geometry and content of the universe\citep{hu,wmap9year}, 
e.g. the first acoustic peak at $\l\simeq$220 indicating a flat universe. It was the Boomerang balloon experiment in 2000 that firstly shows the first acoustic peak position (in a partial sky) is measured at $\l=197\pm 6$ \citep{boomerang}, later MAXIMA-1 at $\l\simeq220$ \citep{maxima}, and DASI at around 200 \citep{dasi}, VSA at 224 \citep{vsa}, compared to full-sky \wmap measurement at $220.8\pm0.7$ \citep{wmap3yeartem} and the latest \planck result  to be $\l=220.0\pm0.5$ \citep{planckps1,planckps2}. The discrepancy is mainly due to the fact that the acoustic peak positions vary in different parts of the sky. In this paper, we test the statistics of the peak positions from ESA Planck data. This paper is arranged as follows. In Sec.\ref{sec:processing} we describe the data processing of \planck data and we use variance as the statistics on peak positions in Sec.\ref{sec:variance}. We show the excessive peak shifts near the Cold Spot area in Sec.\ref{sec:excessive}. Discussion is in Sec.\ref{sec:discussion}.

%\section{Flat sky approximation} 
%The signal on the sky can be analyzed using standard spherical harmonic analysis. For small square patches, however, one can use Fourier analysis: 
%\begin{equation}
%T({\bi r})=\sum_{\bi k} a_{\bi k} \exp\left[\frac{2\pi i( {\bi r} \cdot {\bi k})}{N}\right],
%\end{equation}
%where ${\bi r}\equiv(\theta, \varphi)$ and ${\bi k}\equiv (\kt, \kp)$ if one choose the patch %parallel to the spherical coordinates on the equator. 
%The scaling relation between integer Fourier wavenumber $k\equiv |{\bi k}|$ and multipole number %$\l$ is $\l=2 \pi k /L$, where $L$ is the patch size, so the largest-scale power one can get is at %$\l_{\rm min}=2\pi/L$\footnote{Usually one associates multipole number $\l$ to a characteristic %angular scale $\varpi$ on the sphere via $\l=\pi/\varpi$ because a characteristic angular scale %(e.g. a scale between a cold and a hot spot) is half of one full wavelength, i.e. $\varpi=L/2$.}.
%The angular power spectrum $\cl$ at multipole number $\l$ is scaled from $C_k\equiv \langle|a_{\bi 
%k}|^2 \rangle$ at Fourier wavenumber $k$ via
%\begin{equation}
%C_{\l= 2\pi k/L}=  L^2 C_k,
%\end{equation}
%where the angle brackets denote average over all $ |a_{\bi k}|^2$ for $k-1/2 \le |{\bi k}| < %k+1/2$. Note that the scaling relation can be applied with minimum error for patches centred at %$\theta=\pi/2$ if one uses non-equal area pixelization scheme. Here we use GLESP %scheme\citep{glesp}. The scaling relation is tested in  \citet{direct}. 

\section{Planck data processing and fitting}\label{sec:processing}
To test the statistics of the acoustic peaks, we use flat sky approximation (FSA) and take patches of 20$\times$20 deg$^2$ from the 4 \planck CMB maps: SMICA, NILC, SEVEM and Commander maps in \glesp pixelization\citep{glesp}, each with a specific foreground cleaning method on either harmonic or pixel domain \citep{planckfg1}. The map is rotated every $22^\circ.5$ (in practice it's the $\alm$ of the map for rotation before being repixelized with \glesp with 2 arcmin/pixel) and we take the patches parallel to the $\theta$ and $\phi$ coordinate lines centered at the equator to minimize the error from pixelization. As such we separate the whole sphere into 90 patches. In our analysis we further use the \planck GAL070 mask \citep{planckfg2} to block areas of heavy foreground contamination around Galactic plane, thus in total there are 45 patches available for test in each map, accounting for 43\% of the whole sky. 

%To test the statistics of the acoustic peaks, we use flat sky approximation on the 4 \planck foreground-cleaned maps are used: SMICA, NILC, SEVEM, and C$-$R maps, each with a specific foreground cleaning method on either harmonic or pixel domain \citep{planckfg}. Patches of 20$\times$20 deg$^2$ are taken from each map. To minimize the errors, the map is rotated every $22^\circ.5$ and the patches are taken at the equator avoiding the Galactic plane. In our analysis we further use the \planck GAL070 mask \citep{planckfg}, thus in total there are 45 patches, accounting for 43\% of the whole sky (see Figure \ref{patchtaking}). The re-scaling relation and simulation testing from Fourier to spherical harmonic analysis is shown in \citet{direct}. 

\begin{figure*}
\begin{center}
\includegraphics[width=0.35\textwidth]{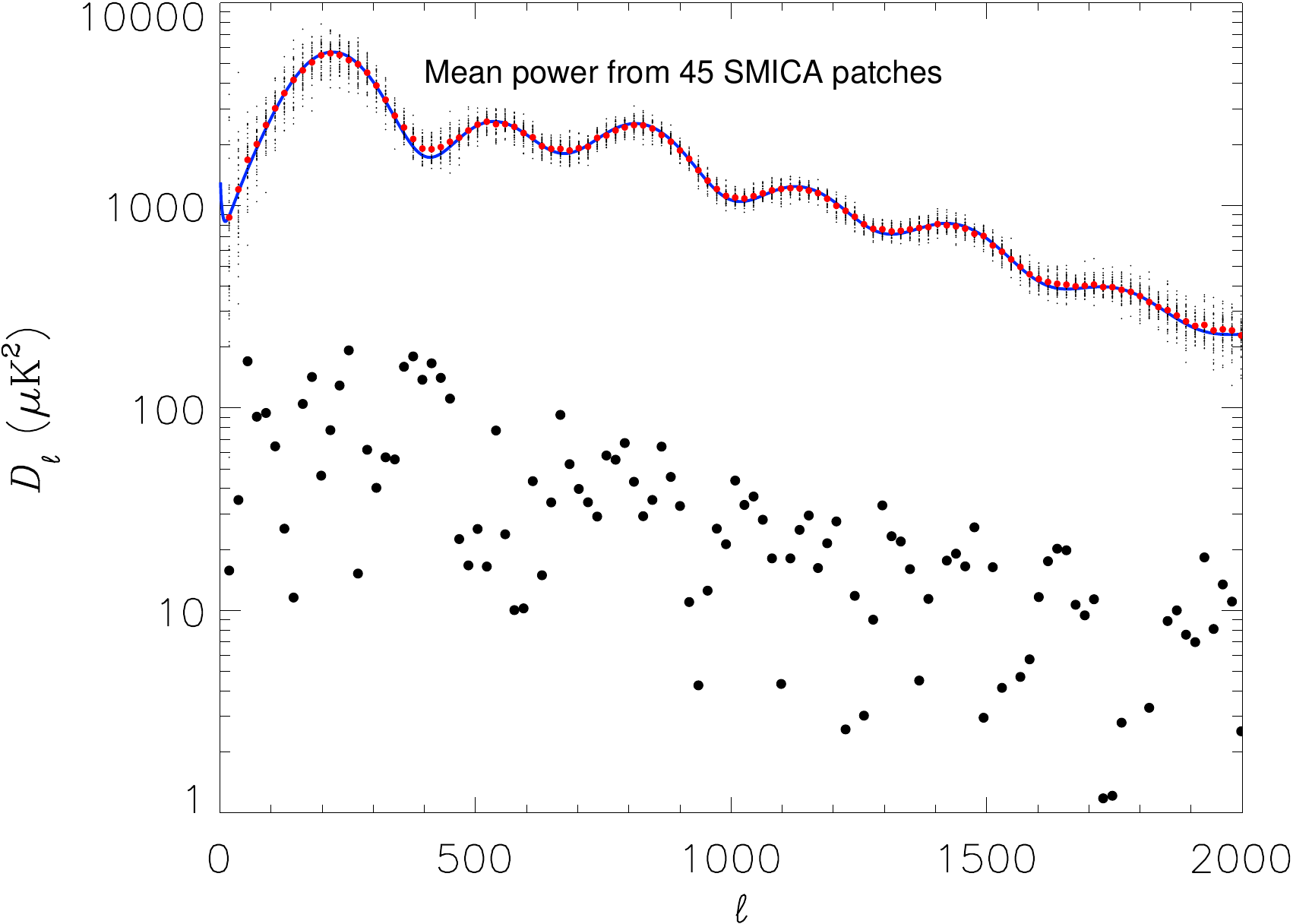}
\includegraphics[width=0.35\textwidth]{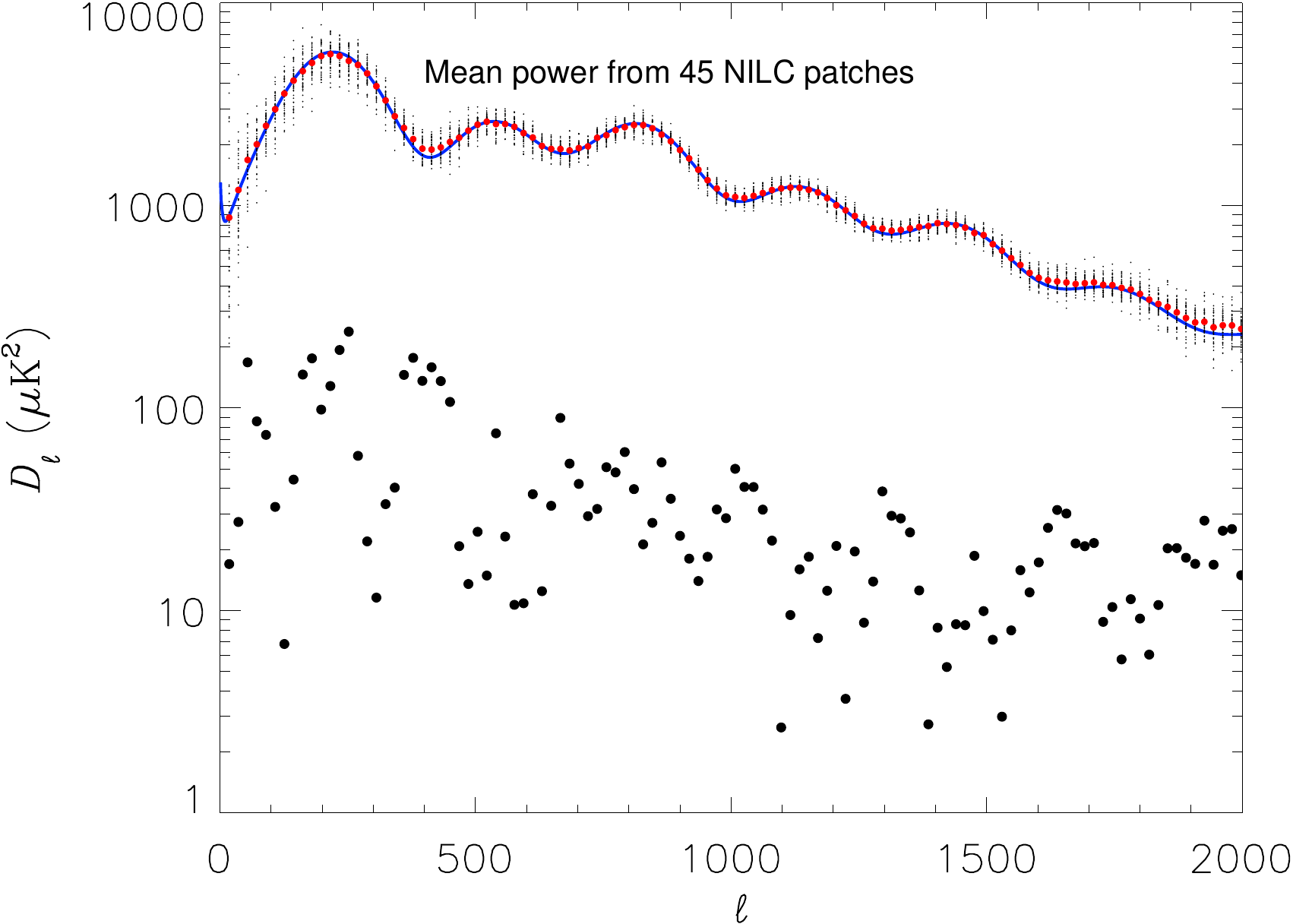}\\
\includegraphics[width=0.35\textwidth]{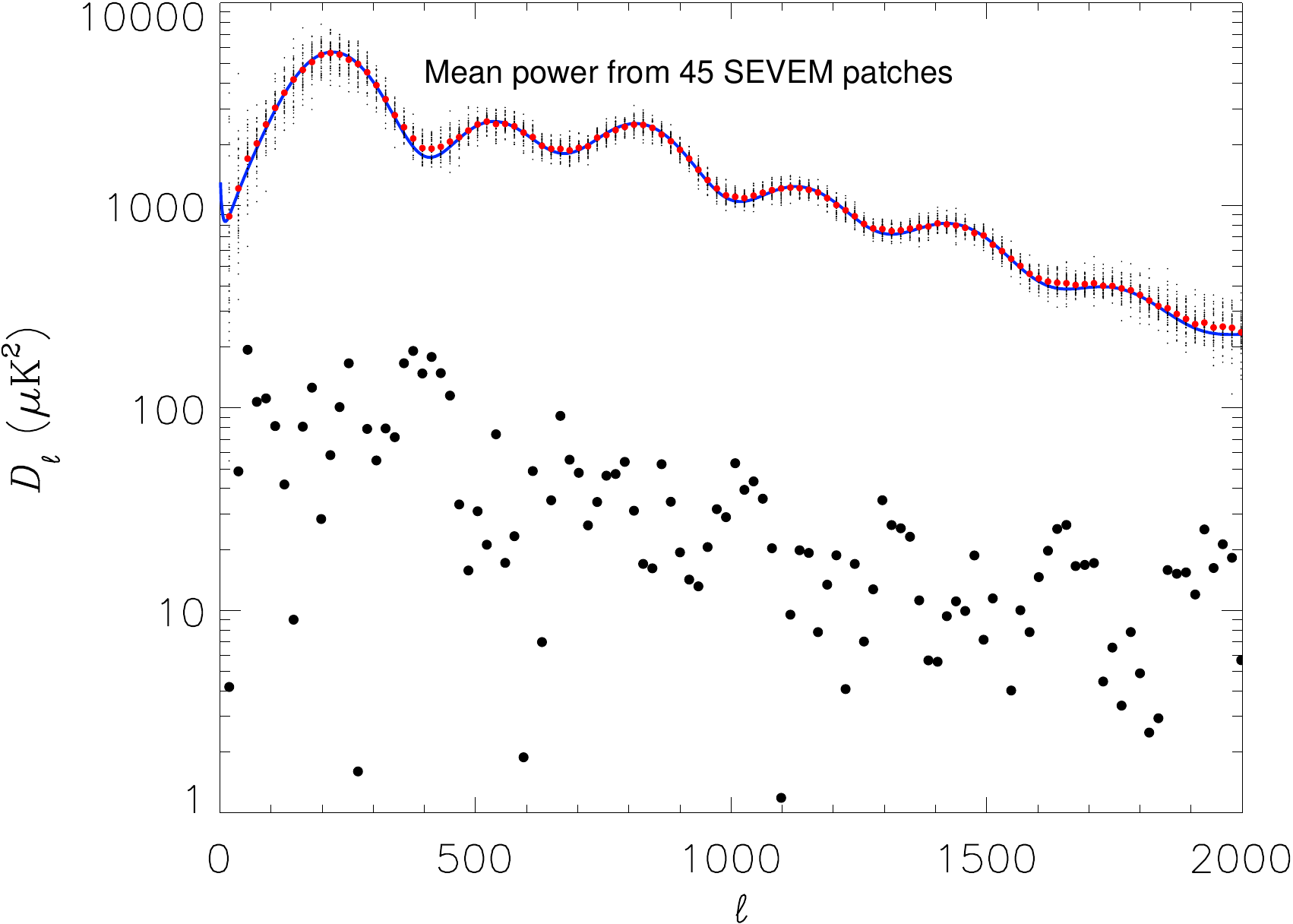}
\includegraphics[width=0.35\textwidth]{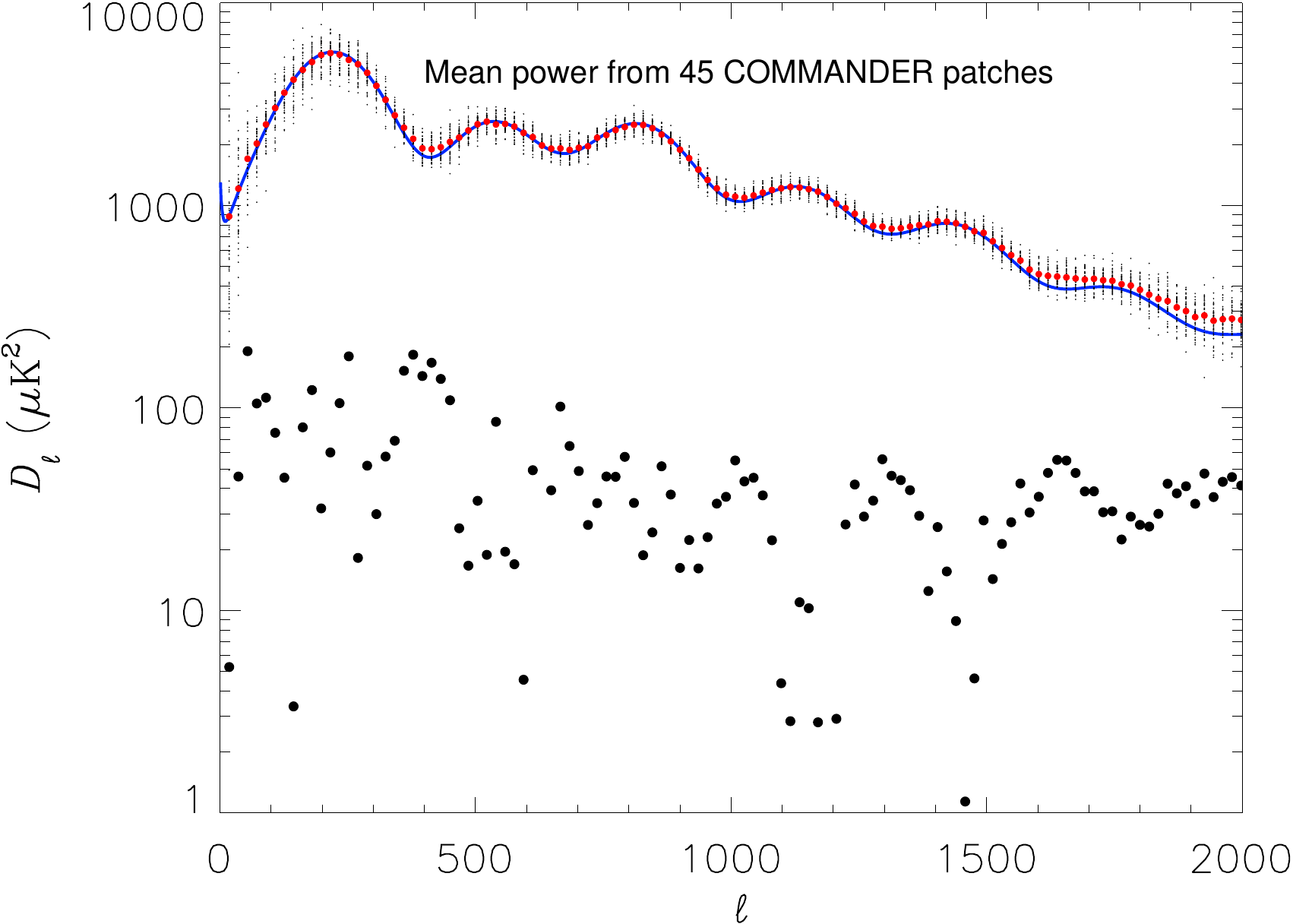}\\
\caption{The angular power spectra from 45 patches of the \planck 4 maps. Top part in each panel is the mean power spectrum (big red dot) from the 45 patches (small black dot) and the \planck best-fit $\Lambda$CDM power spectrum (blue curve), whereas the bottom part (big black dot) is the absolute difference between the mean and the \planck best-fit spectrum.} 
\label{dl}
\end{center}
\end{figure*}

Each patch is firstly interpolated in the $\theta$ direction as the spacing of the grid is slightly uneven due to the fact that the grids are where the Gauss-Legendre polynomial zeros are \citep{glesp}. Two-dimensional Fourier transform is used to analyze each patch. %\footnote{We choose FSA over any full-sky power spectrum code that can deal with partial sky because the latter usually comes with apodization process which significantly modifies the power spectrum.} 
The scaling relation between integer Fourier wavenumber $k$ and multipole number $\l$ is  $\l=2 \pi k /L$, where $L$ is the size of the patch. The amplitude of the angular power spectrum $S_\l$ can be scaled from $S_k$ at Fourier wavenumber $k$ via
\begin{equation}
S_{\l\equiv 2\pi k/L}=  L^2 S_k.
\end{equation}
So for each patch with size $L$, the wavenumber $k$ is scaled into $2\pi k/L$, and there is no binning in our power spectrum and the sequence of $k$ is scaled such that the multipole interval $\Delta \l\equiv 2\pi/L$.
One can also see the choice of the patch size cannot be arbitrary. The sampling interval $\Delta \l$ of angular power spectrum is inversely proportional to $L$, so we want to choose patches as large as possible to have more sampled data in the power spectrum, but it can't be too large as the curvature of the sphere causes the flat sky approximation to break down. The scaling relation from Fourier to spherical harmonic analysis is already tested with simulations \citep{direct}. 

One issue related to Fourier transform on real signals is the non-periodic boundary (NPB) condition, which induces extra power from jump discontinuity. The condition is aggravated when there is beam convolution causing jump discontinuity more pronounced at the boundary. We have conducted simulations to estimate the extra power due to 5-arcmin beam convolution on a 20$\times$20 deg$^2$ patch to be $1.21\, \ell^{-1.38}$ ($\mu$K$^2$) (see the Appendix A). This should be subtracted before any data processing.

It is assumed that the 4 \planck maps are fairly cleaned such that the angular power spectrum in each patch $S_k=b_k^2 C_k +N_k$,  where $C_k$ denotes the CMB, $b_k$ the beam transfer function and $N_k$ the noise. We then use cross-power spectrum (XPS) to reduce the noise (see Appendix B). The XPS is a quadratic estimator between two patches or maps to provide an unbiased estimate of the power spectrum of the correlated signals: 
\begin{equation}
b^2_kC_k^{\alpha\beta}= \frac{1}{2} \langle  \alpha^{*}_{\bi k} \beta_{\bi k} + \beta^{*}_{\bi k} \alpha_{\bi k} \rangle,
\label{def}
\end{equation}
where  $\alpha_{\bi k} $ and $\beta_{\bi k}$ are the Fourier modes of patches from \planck halfring maps, $*$ denotes complex conjugate and the angle brackets denote an average over all $k-1/2 \le |{\bi k}| < k+1/2$ for integer k. The residual of the XPS $X_k^{\alpha\beta}$ has to be further subtracted, which stems from the lack of an ensemble and non-zero chance correlation between the "uncorrelated" signal  \citep{xps,direct}, 
\begin{equation}
\sqrt{\langle (\xk^{\alpha\beta})^2 \rangle} \simeq \frac{\sqrt{ N^\alpha_k N^\beta_k } }{\sqrt{2\pi k}},
\label{xpsuncorr}
\end{equation}
where $N^\alpha_k$ and $N^\beta_k$ are the (uncorrelated) noise power spectrum from halfring patches, respectively. The noise power spectrum can be obtain by $1/2$ of the differencing of halfring patches, assuming both are of the same level. Thus in our data pipeline, we subtract the residual $X_k^{\alpha\beta}$ from the XPS, following the correction of NPB.

\begin{figure}
\begin{center}
\includegraphics[width=0.45\textwidth]{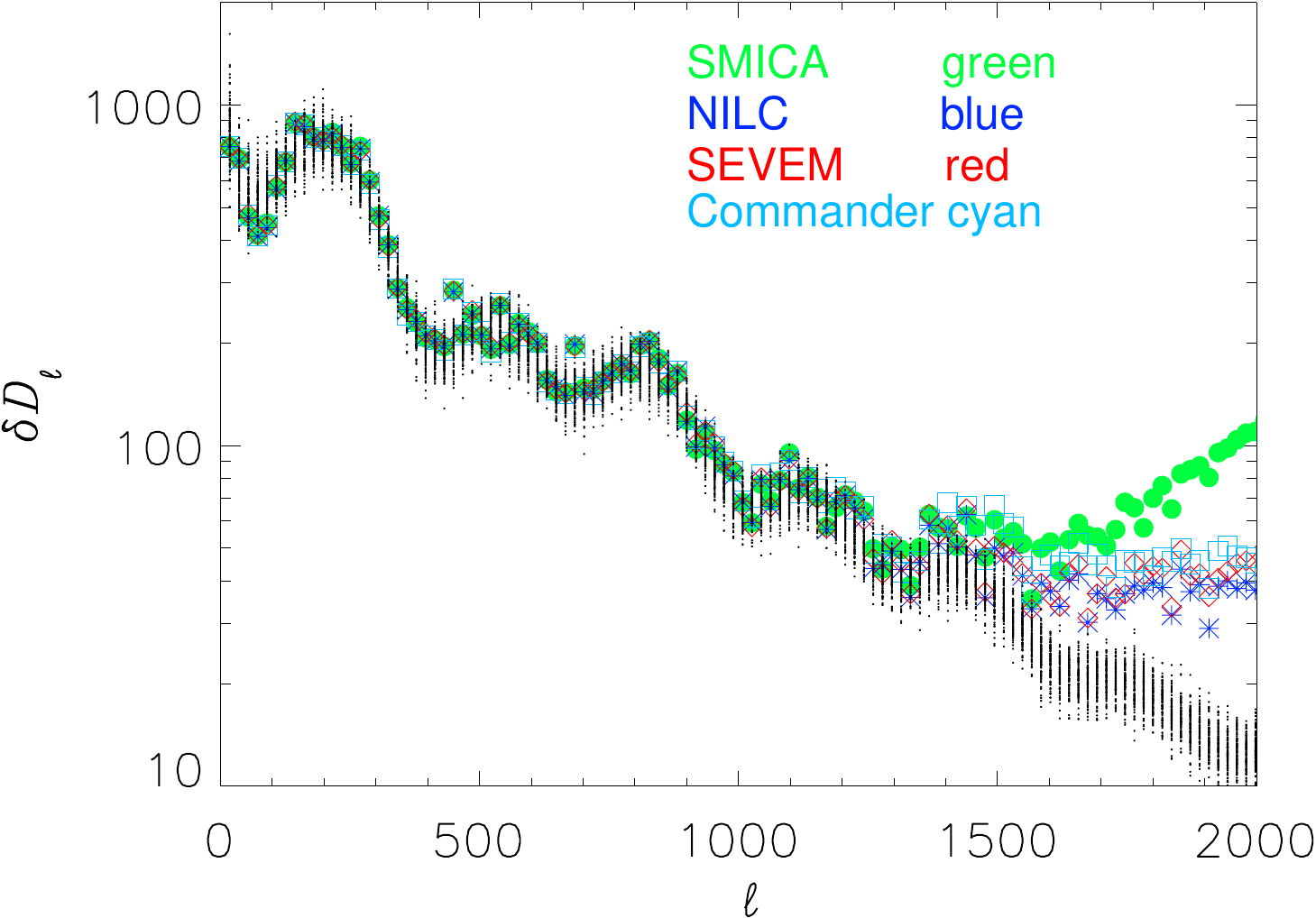}
\caption{The dispersion of the 45 angular power spectra of the Planck patches with green filled circle, blue asterisk, red diamond and and light blue square sign denoting SMICA, NILC, SEVEM and Commander patches, respectively and those from the 100 simulations (black dot). One can see the dispersions from the Planck 4 maps fall within the 100 simulations up to $\l= 1,350$, which is then adopted as the the upper limit of our Gaussian-fitting range.} 
\label{dispersion}
\end{center}
\end{figure}

It is then subsequently de-convolved with a beam transfer function at 5 arcmin \citep{planckfg2}. To check the validity of our data processing, in the 4 panels of Figure \ref{dl} we show the mean angular power spectrum from the 45 patches of the \planck 4 maps, which fits very well with the best-fit $\Lambda$CDM power spectrum. And in Fig \ref{dispersion} we show the dispersions of the 45 power spectra from the \planck patches, which fall within the range of the simulation up to $\l=1,350$.

%It is assumed that the 4 \planck maps are fairly cleaned such that in each patch the signal $S= C\otimes B+N$ where $C$ denotes the CMB, $B$ the effective beam, $N$ the noise and $\otimes$ convolution. Each patch is de-noised by subtracting the mean power from $\l$=2200 to 2450 where the noise dominates the power spectrum, and then subsequently de-convolved with the window function provided in \citet{planckfg}.  

%\begin{figure}
%\centering
%\epsfig{file=gal070_patch.eps,width=8cm}
%\caption{The $20\times20$ deg$^2$ patches with \planck GAL070 %mask.}
%\label{patchtaking}
%\end{figure}

\begin{figure}[ht]
\begin{center}
\includegraphics[width=0.5\textwidth]{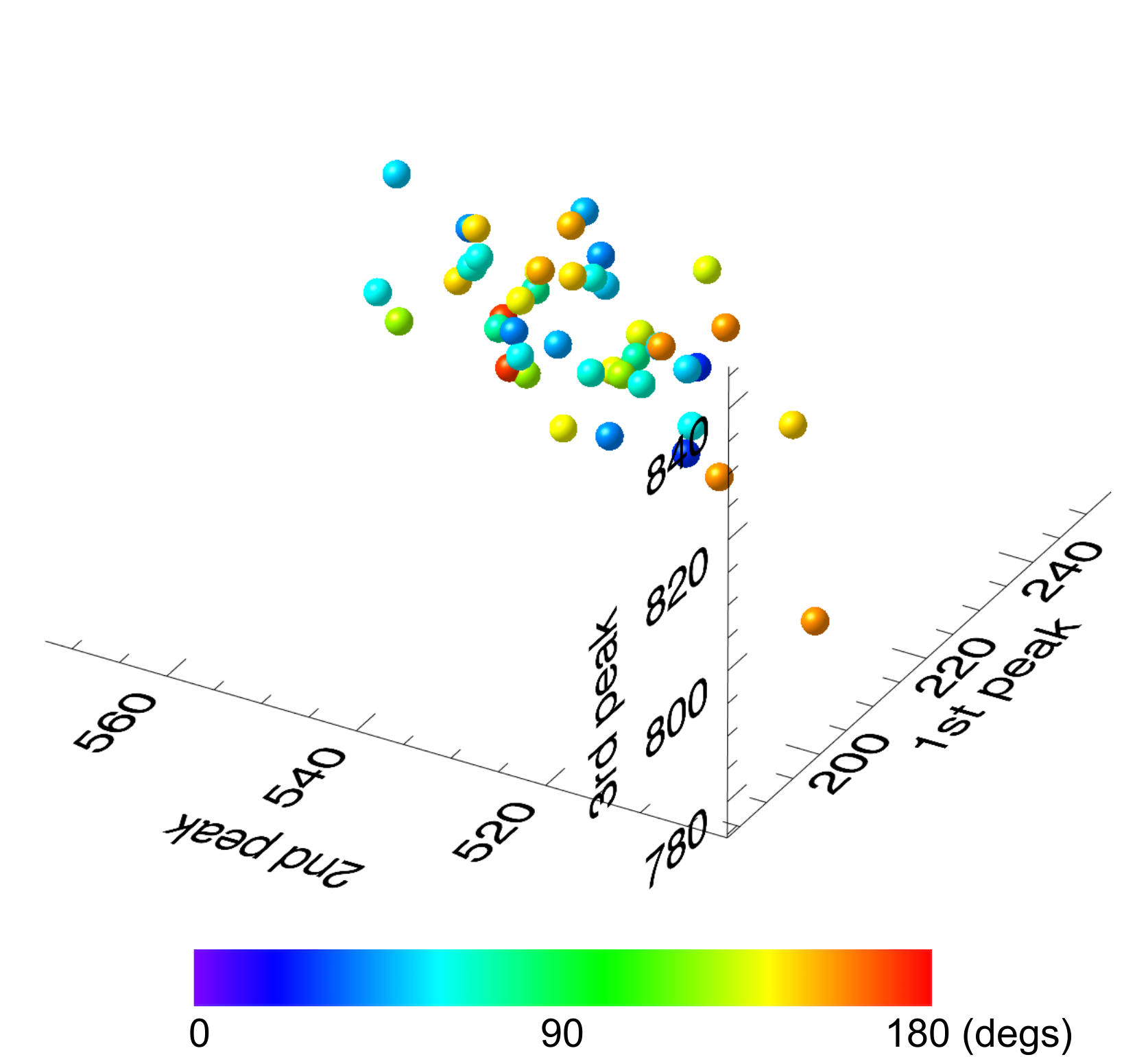}
\includegraphics[width=0.45\textwidth]{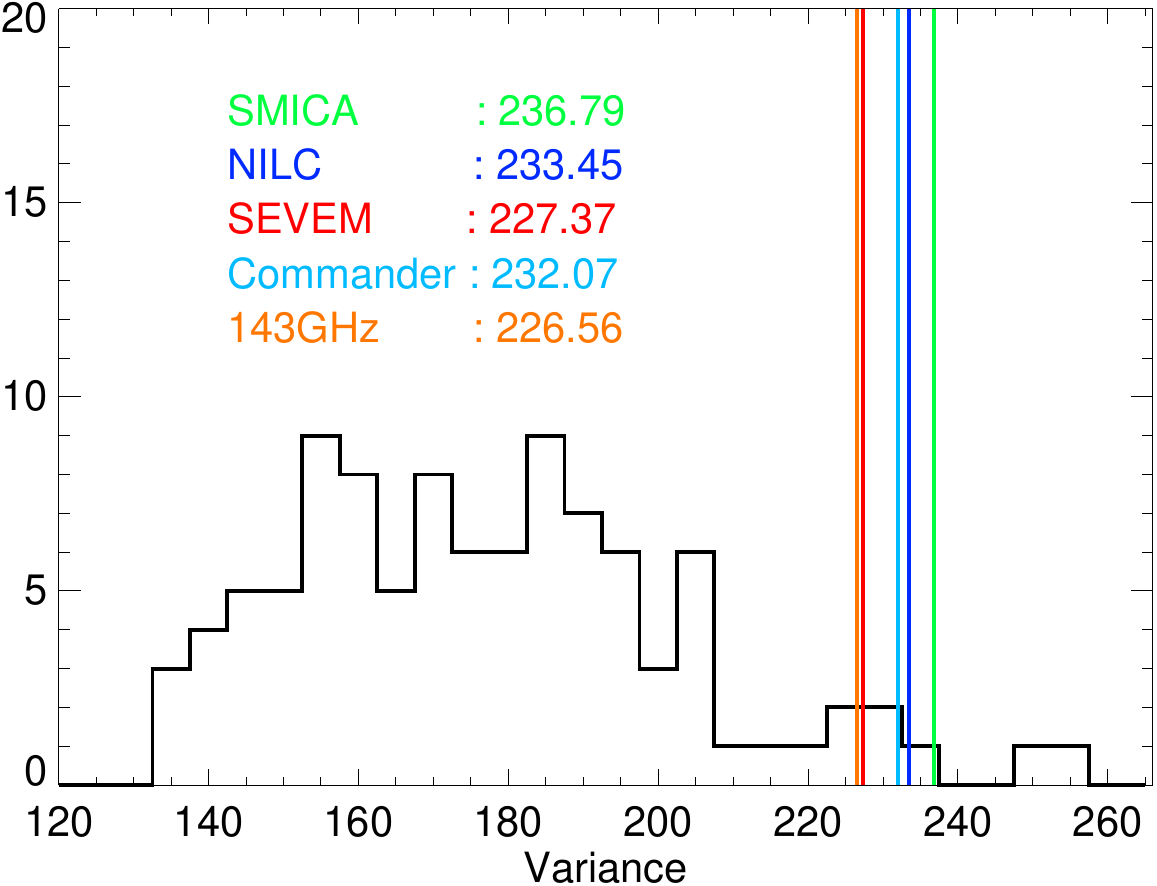}
\caption{Distribution of the acoustic peak positions of the SMICA 45 patches and histogram of the variances of the peak positions from 100 realizations of \planck simulation. In the top panel each patch is represented with one point ${\bi \l}\equiv(\l^{(1)}, \l^{(2)},\l^{(3)})$ with the peak positions denoted in 3 coordinates. The color in each point denotes the angle between the Galactic North Pole and the patch center. In the bottom the variances of the \planck 4 maps and 143 GHz band map are denoted with colored lines.} 
\label{variance}
\end{center}
\end{figure}

We use the Levenberg-Marquardt algorithm \citep{recipes} as the non-linear least-square fitting routine to fit the angular power spectrum curve with sum of 5 Gaussian functions to the beginning of the 5th peak (from $\l=50$ to 1,350). The peak positions are estimated in $\l$ and only the first 3 peaks are estimated because the low amplitudes of the 4th and 5th peaks are prone to error in the estimation. The distribution of the peak positions of the SMICA map is represented in 3 dimensions of ${\bi \l}\equiv(\l^{(1)}, \l^{(2)},\l^{(3)})$ in top panel of Figure \ref{variance}.

%We use 5 Gaussian functions and the Levenberg-Marquardt algorithm as the non-linear least-square fitting routine to fit the power spectrum curve to the beginning of the 5th peak (up to $\ell=1314$) \citep{recipes}. Sinusodial functions are also used to fit, but simulations shows fitting with 5 Gaussian functions has both more stable fitting and better accuracy. 

Any statistics from the data are to be compared with lensing simulations by \planck Consortium \citep{plancklensing1}. There are 100 realizations based on best-fit $\Lambda$CDM model with both the lensing and the Doppler boosting effects included \footnote{\tt{http://wiki.cosmos.esa.int/planckpla/index.php/Simulation\_data}}, which are originally employed to determine the lensing mean field, the measurement error bars, and to validate lensing reconstruction methodology. The angular resolution of the simulations is 5 arcmin, the same as the 4 Planck maps. Gravitational lensing effect \citep{blanchard,cole,linder,seljak,challinorlense,lewis}
is the deflection of CMB photons coming from the last scattering surface by potential gradients along our line of sight, and the \planck team used both temperature and polarization data to measure the lensing potential at a level of 40$\sigma$ \citep{plancklensing1,plancklensing2}. The Doppler boosting effect \citep{challinorpeculiar}, on the other hand, is due to the motion of our solar system barycenter with respect to the CMB rest frame that induces not only the dipolar anisotropy, but also the Doppler modulation and aberration at high multipoles \citep{planckdb}. Both effects on the angular power spectrum at high multipoles are small, but they can affect significantly the angular power spectrum of a small patch \citep{jeong}.  Since both effects are still in the \planck 4 maps, it is thus appropriate to take for comparison the simulation with the aforementioned  effects included. We process the 100 simulations with the same pipeline as the \planck 4 maps with the same choice of patches and rescaling.

\section{Variance of the peak positions}\label{sec:variance}
We use the variance of the first 3 peaks to characterize the statistics of peak positions:
\begin{equation}
\sum^n_{i=1} |{\bi \l_i}-\bar{\bi \l}|^2/(n-1), 
\end{equation}
where $\bar{\bi \l}$ is the mean of the peak positions from the n=45 patches. In bottom panel of Figure \ref{variance} we plot the histogram of the variance from 100 CMB realizations and those of the 4 \planck maps are denoted with colored lines. The result shows that the variances of the peak positions from 4 maps are all quite large with significance $p< 0.04$, in particular the variance of the SMICA map is higher than 98 realizations, which is equivalent roughly to a significance $p$-value 0.02 against the null hypothesis. 

%In Figure \ref{result} we plot the histograms of the variance from 100 CMB realizations for the 1st acoustic peak (top left), the principal component (top right) and the total variance (bottom left). The result shows that the variances of the peak positions from 4 Planck maps are all higher than usual. In particular for the total variance only 1 out of 100 CMB realizations is higher than the SMICA and NILC maps, which is equivalent roughly to a significance $p$-value 0.01 against the null hypothesis. It is observationally evident that cosmological CMB photons dominate at high Galactic latitudes, thus to further reduce the effect of Galactic foreground residual, we use \planck GAL060 mask (38 patches) for total variance and SMICA map is higher than all 100 realizations, pushing $p<0.01$. 

We examine the possible effect from foreground residual with the mean and the dispersion of the 45 power spectra from the patches. It is based on the fact that any extra non-correlated signal added to the CMB would increase both the mean and dispersion of the power spectra. As shown in Figure \ref{dl} the mean angular power spectra fit well with the best-fit $\Lambda$CDM power spectrum, and in Figure \ref{dispersion} the dispersions from the 45 patches fall within the 100 realizations up to $\l= 1,350$. Thus we are guaranteed that the data is at least free from significant foreground residual up to $\l=1,350$.

%\section{beam}
%We examine the systematic effect by non-symmetric effective beam. We simulate Planck SMICA map with FWHM 4.85 arcmin with effective beam axis ratio 1.13. Here we assume a slow rotating scenario such that the angle of the effective beam axis is kept fixed within 20$\times$20 deg$^2$ patches. The CMB patch is convolved with the effective beam with 9 different angles (20 degrees apart), from which we calculate the root-mean-square of the positions against the average, and the largest r.m.s is 0.64 among 25 sets of convolved patches, a strong indication that non-symmetric beams with different angles is not the main cause of the anomaly.

%\section{foreground residual}

If foreground residual is the cause of the anomaly, the variance from any frequency band map should be higher than that of the SMICA map because the foreground contamination in any band map is definitely higher than that in SMICA map. However, the variance for the \planck 143 GHz channel map containing diffuse foreground is 226.56 (where we only subtract bright point sources), smaller than that of SMICA map 236.79 instead (where the distance between the mean peak positions of 143GHz and those of SMICA is only  2.19). The foreground residual as the cause for the anomaly can then be ruled out. The influence of the Galactic plane is also tested by color-coding the points for the distribution of the peaks. There is no obvious clustering in color, hence no systematic effect from either Galactic (as shown in top panel of Figure \ref{variance}) or Ecliptic influence.

\begin{figure}[ht]
\begin{center}
\includegraphics[width=0.4\textwidth]{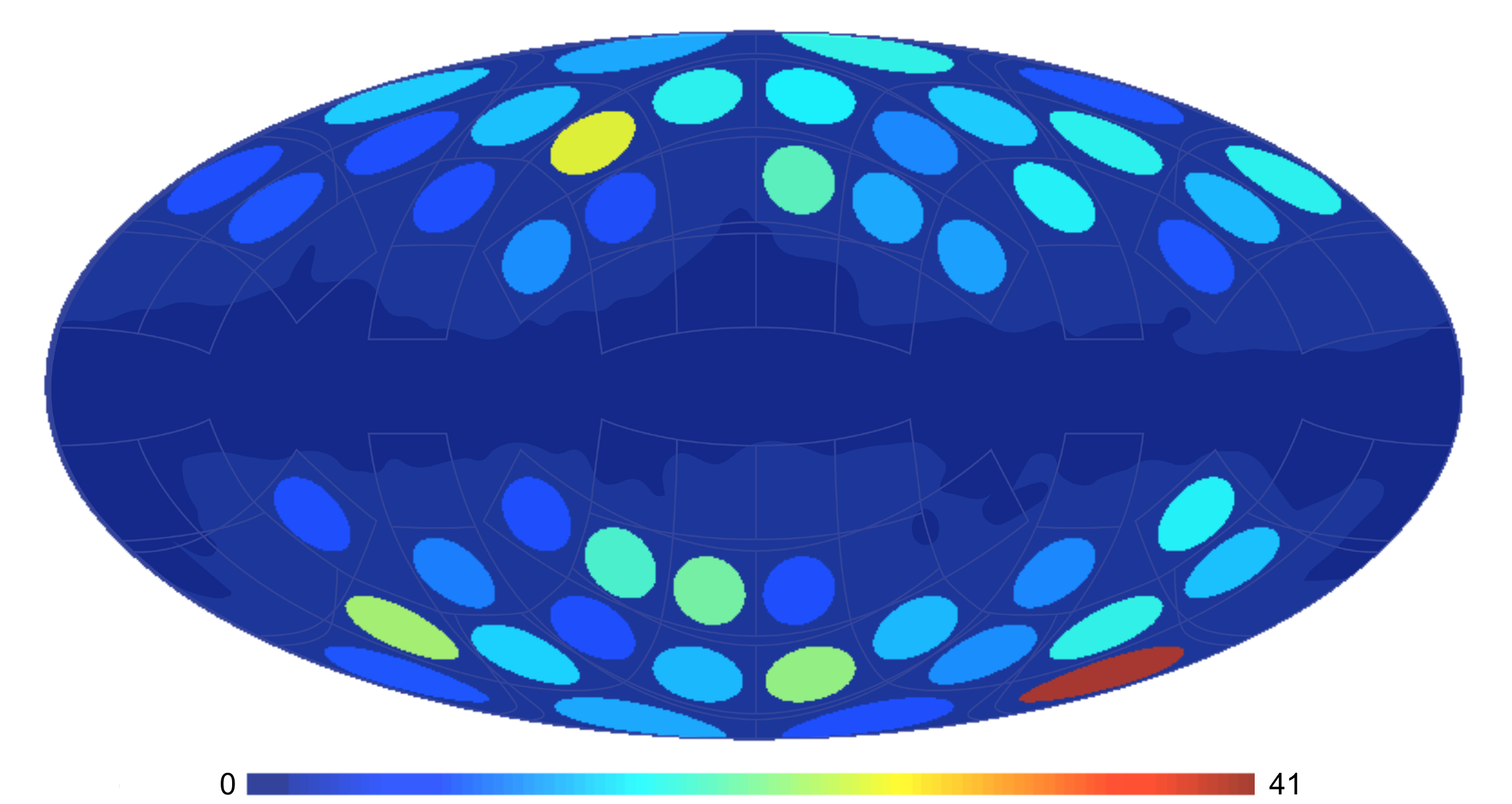}\\
\includegraphics[width=0.3\textwidth]{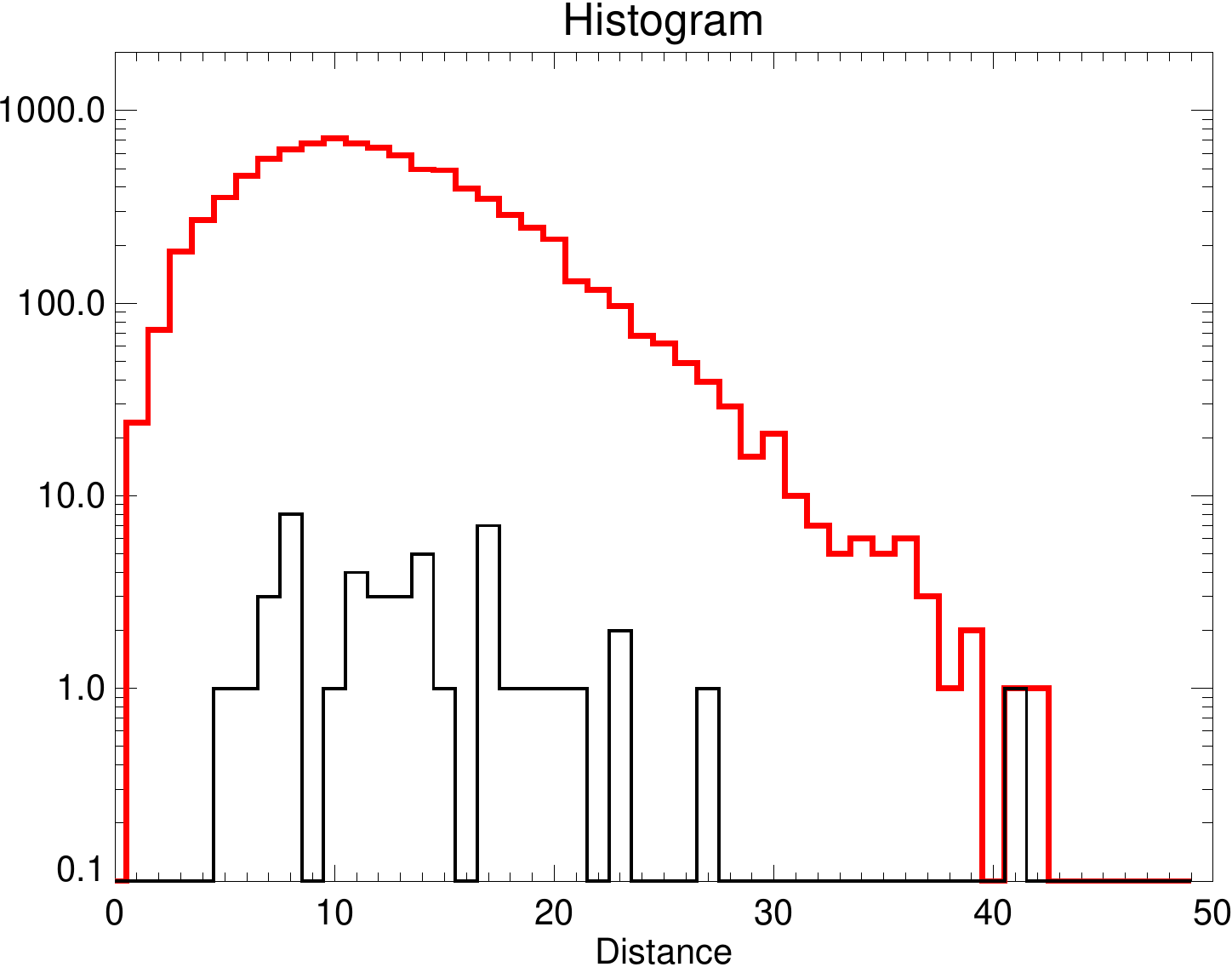}
\caption{The peak distance from the mean: $|{\bi \l_i}-\bar{\bi \l}|$ of the 45 patches on the sky (top), and their histogram (black curve in the bottom panel). In the top panel we use large round symbol centered in each patch to denote the distance with color, and the patch with red color is centered at $(l,b)=(204^\circ.74,-65^\circ.48)$, where the famed Cold Spot is inside. In the bottom panel the red curve is the histogram of the 9000 patches from the \planck simulation.}
\label{distance}
\end{center}
\end{figure}
% OK %%%%%%%%%%%%%%%%%%%%%%%%%%%%%%%%%%%%%%%%%%%%%%%%%%%%%%%%%%%%%%%%%%%%%%%%%%%%%%%%%%%%%%%%%%%%%%  

\section{Excessive peak shifts in the Cold Spot patch}\label{sec:excessive}
\begin{figure}
\begin{center}
\includegraphics[width=0.2\textwidth]{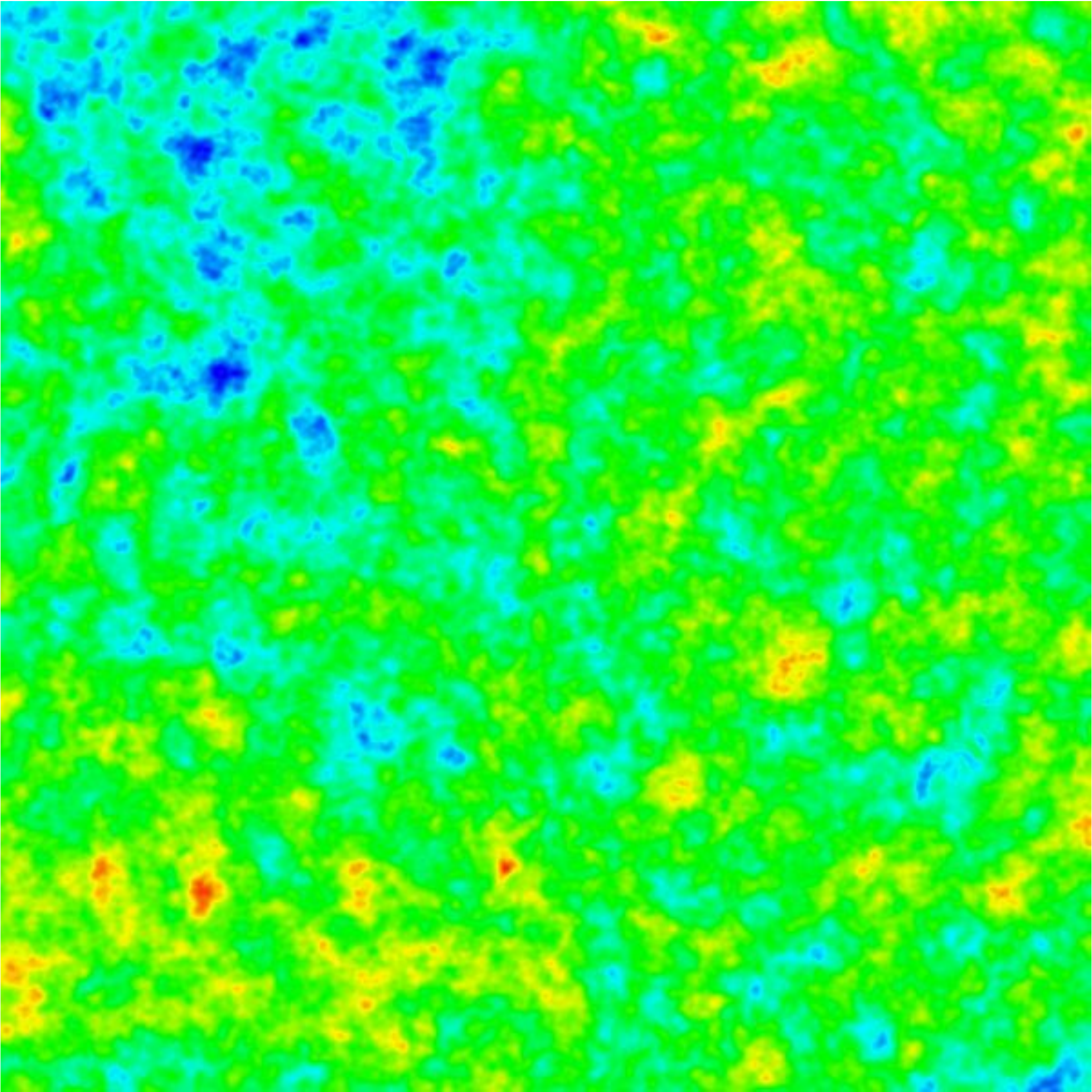}
\includegraphics[width=0.2\textwidth]{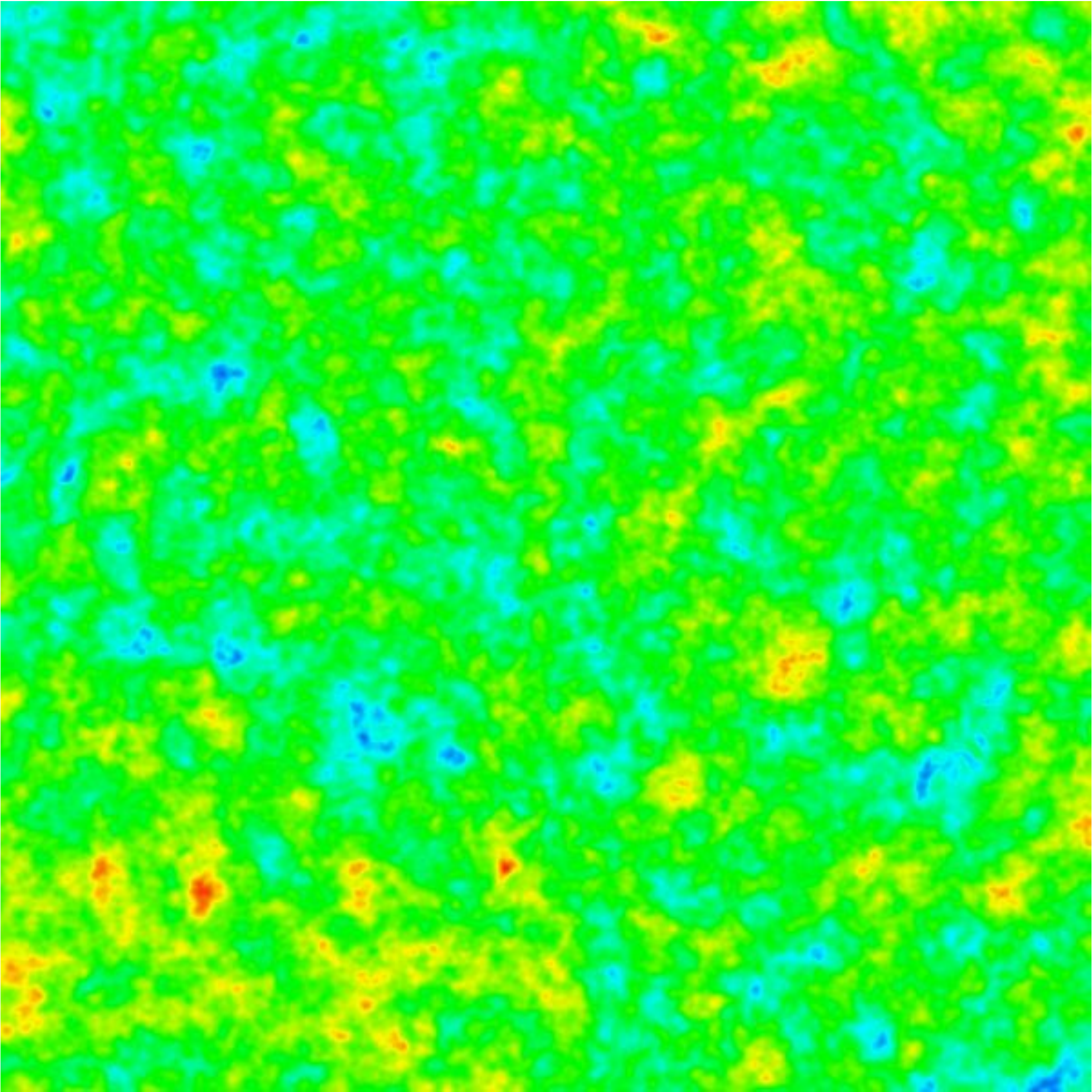}
\includegraphics[width=0.4\textwidth]{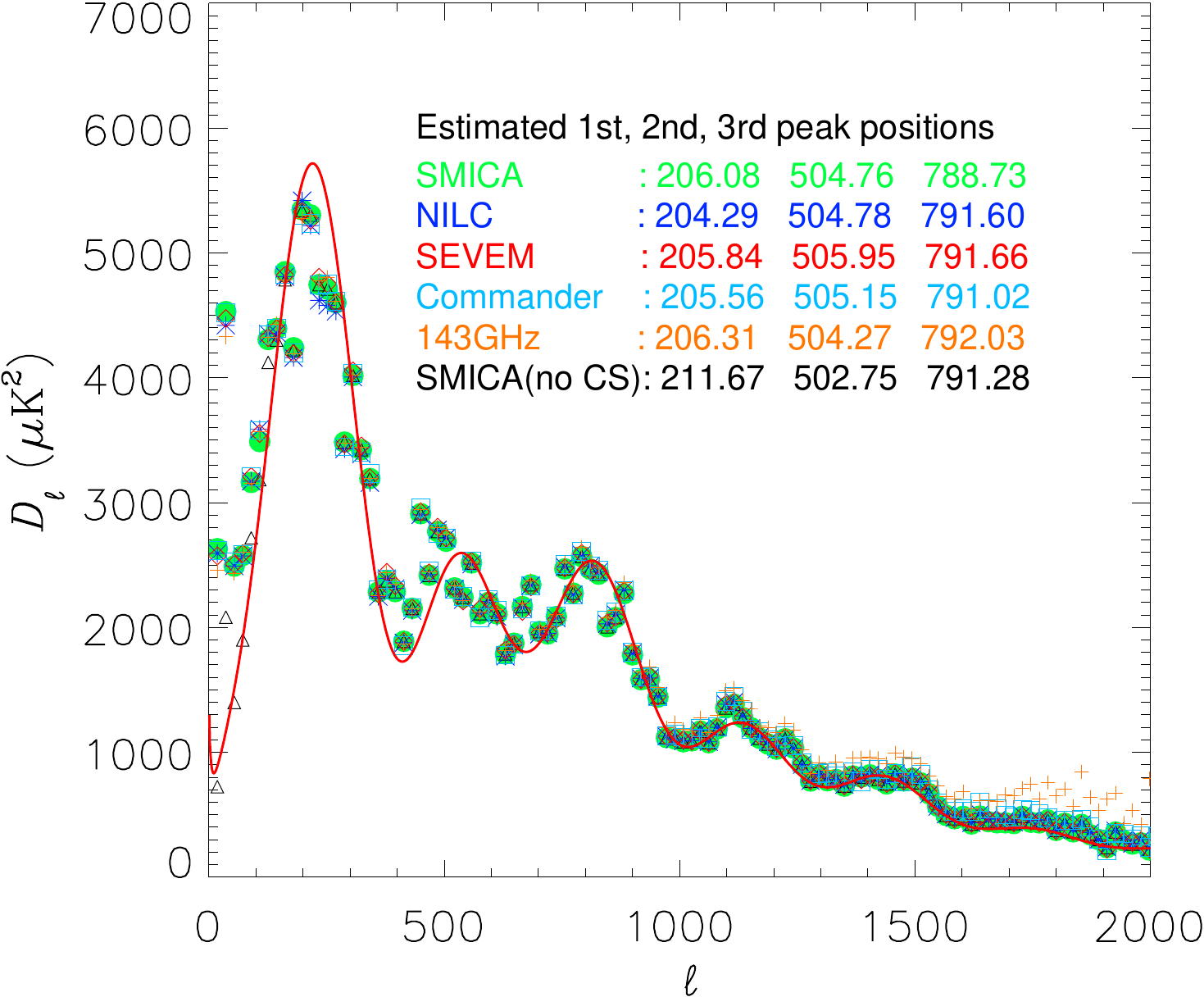}
\includegraphics[width=0.4\textwidth]{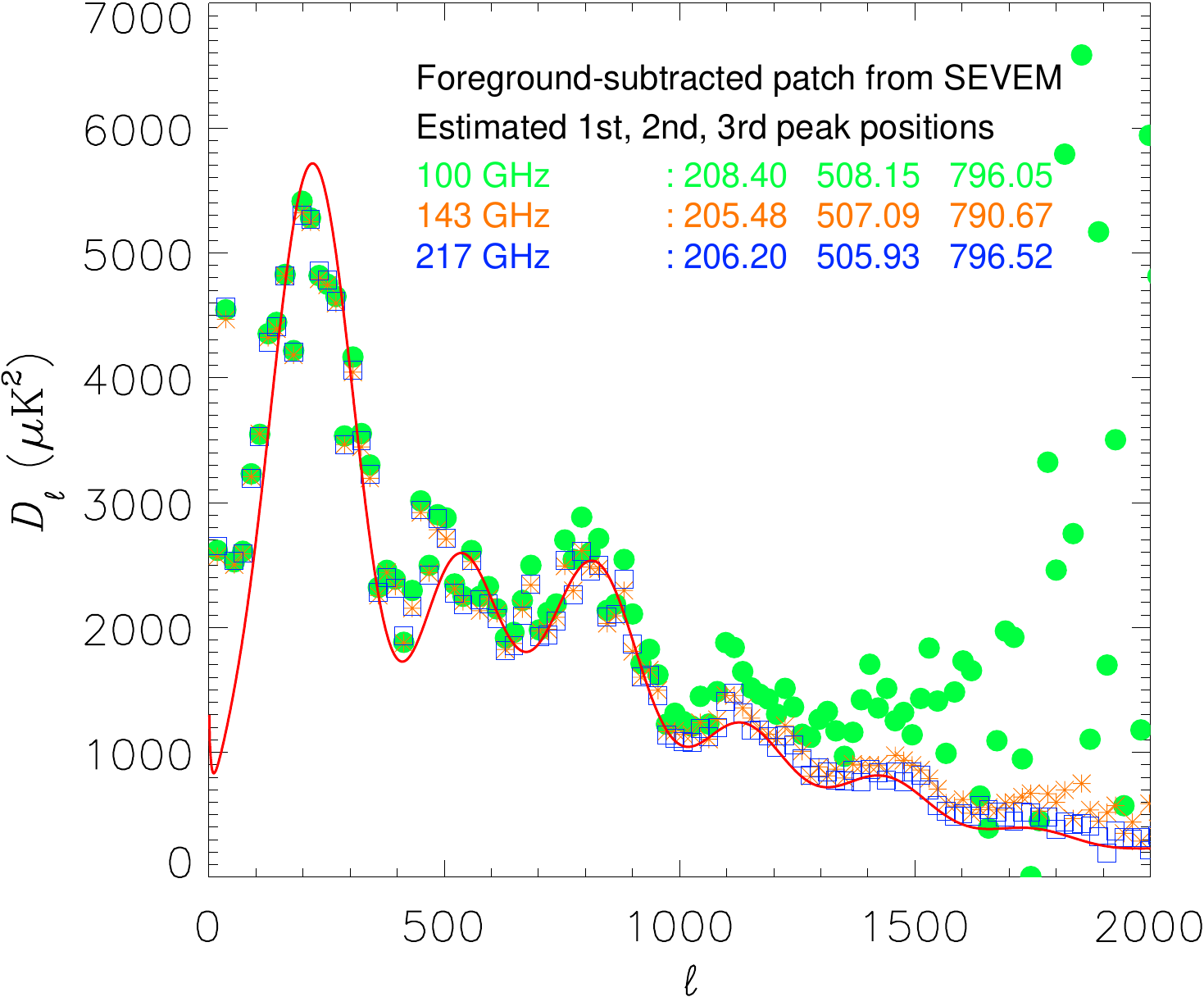}
\caption{The patch containing the well-known CS from SMICA map (top left panel) and with the  CS subtracted (top right), where color bar range is $[-500,500]\mu$K. Middle panel shows the angular power spectra of the CS patch from \planck 4 maps with green filled circle, blue asterisk, red diamond and light blue square sign denoting SMICA, NILC, SEVEM and Commander patches, respectively. One can see the data points of the first 3 peaks are mostly on the left side of the best-fit $\Lambda$CDM model (red curve). For comparison, we also show that of the same field from \planck 143 GHz map (orange plus sign), which still contains foreground, and is corrected only for the extra power from non-periodic boundary condition. Even with foreground, the peak positions at 143GHz are close to those of the CMB patches. Also shown is that of the CS-subtracted square (black triangle). In the bottom panel, we show the same for the CS patch of the foreground-subtracted maps at 100, 143 and 217 GHz from SEVEM method (green filled circle, orange asterisk and blue square, respectively).} 
\label{coldspot}
\end{center}
\end{figure}

Since the variance is related to the "distance" of the peak positions of patch $i$ to their mean, $|{\bi \l_i}-\bar{\bi \l}|\equiv \sqrt{(\l_i^{(1)}-\bar\l^{(1)})^2+(\l_i^{(2)}-\bar\l^{(2)})^2+(\l_i^{(3)}-\bar\l^{(3)})^2}$, we show the distance of the 45 patches on the sky in top panel of Figure \ref{distance}. One can easily notice one patch has far larger distance 40.72 than the rest. That patch is centered at $(l,b)=(204^\circ.74,-65^\circ.48)$ near the Eridanus constellation, which contains the well-known CMB "Cold Spot" (CS) \citep{vielva,cruz} (see top left panel of Figure \ref{coldspot}), an area where the CMB temperature has lower than Gaussian theory predicts with significance around 1\% \citep{planckstat}. One should also note that the 3 peaks of the patch with CS not only have larger shifts than usual, but also display synchronous shifts towards smaller multipole numbers ({\it i.e.} larger scales) than the mean, as shown in middle panel of Figure \ref{coldspot}.

\begin{figure}[ht]
\begin{center}
\includegraphics[width=0.2\textwidth]{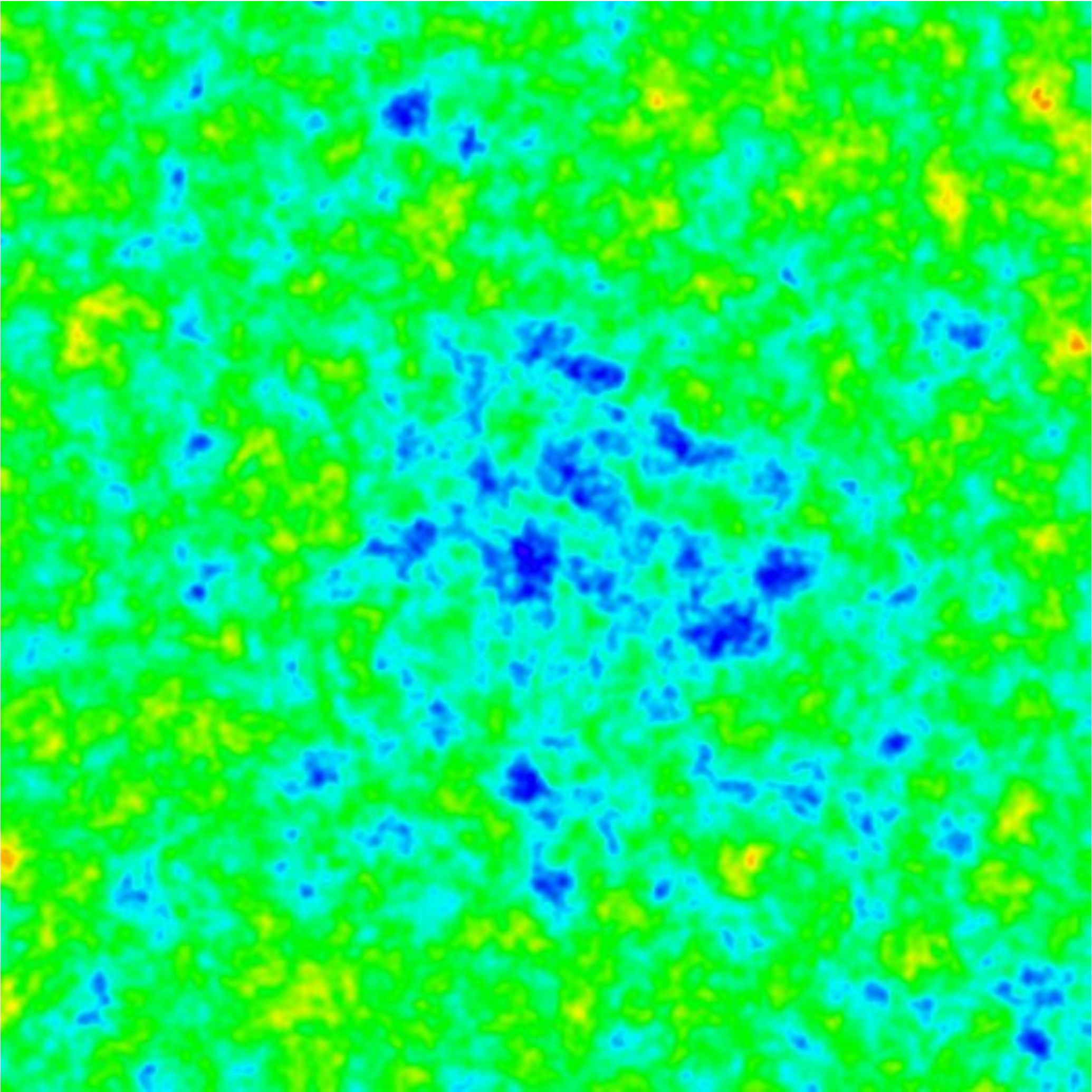}
\includegraphics[width=0.2\textwidth]{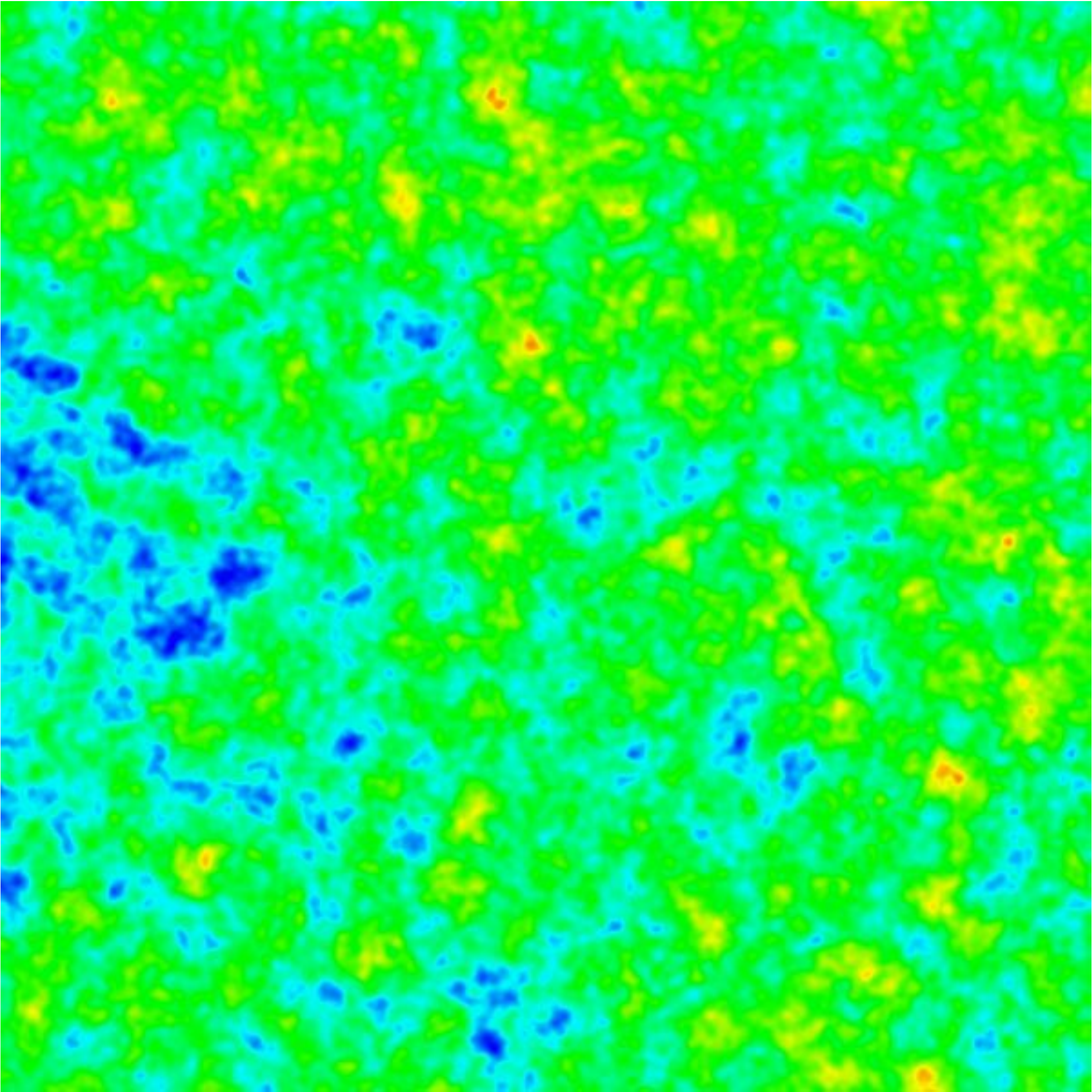}
\includegraphics[width=0.47\textwidth]{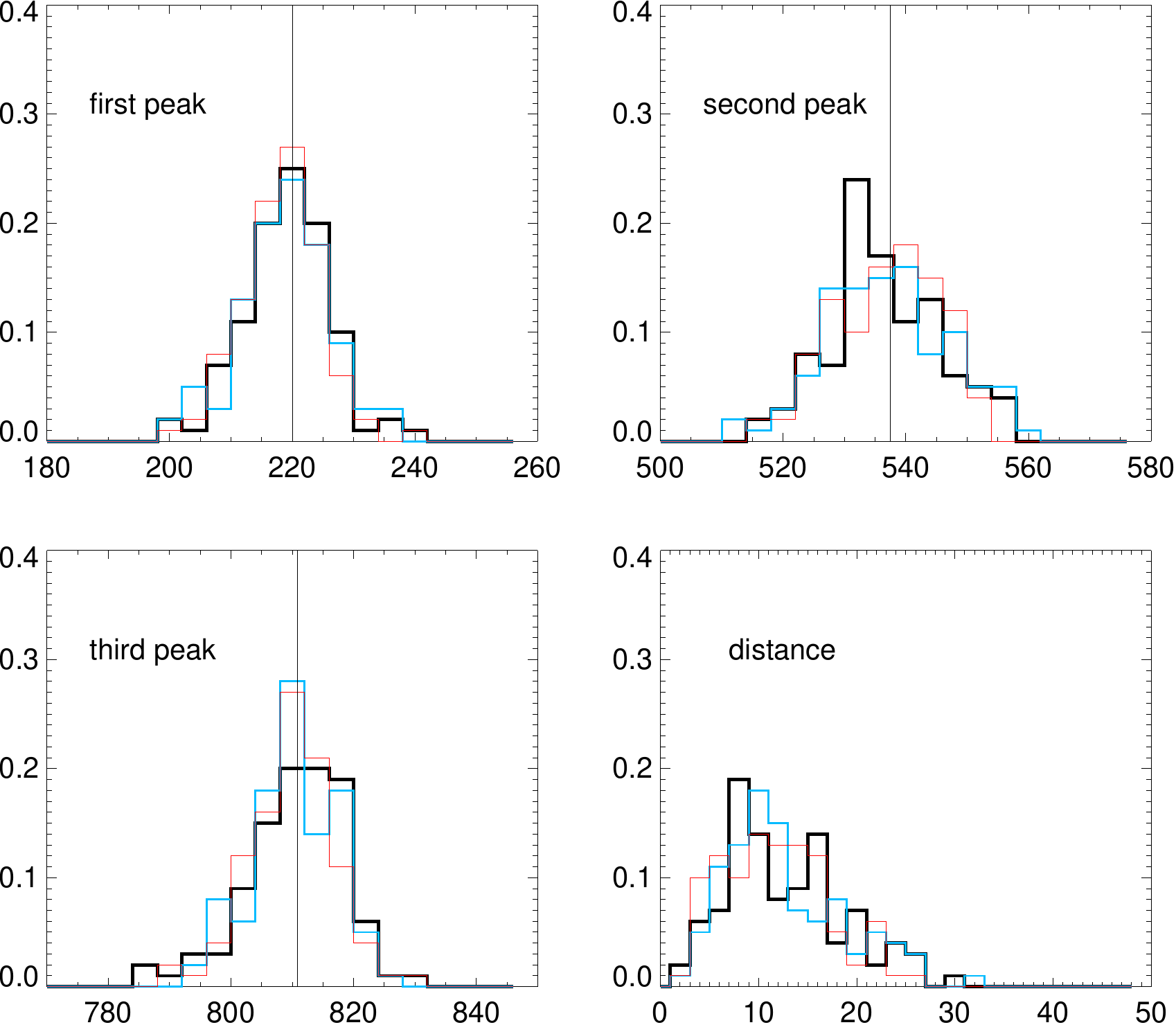}
\caption{Top left panel is an example of a simulated cold spot placed roughly in the center of a $20\times20$ deg$^2$ patch, whereas the right panel is that by shifting 10 deg such that half of the cold spot is on the edge of the patch to mimic the CS patch. Bottom 4 panels show the normalized histograms of the first 3 peak positions and the peak distance from 100 simulated full-cold spots placed at the patch center (black curves) and 100 half-cold spots (blue curves). For comparison, we plot the same for 100 patches taken from the \planck lensing simulation (red curves). The vertical lines denote the \planck best-fit peak positions $ (\ell^{(1)},\ell^{(2)},\ell^{(3)})=(220.0, 537.5, 810.8)$, respectively. One can see the distributions of peak positions and distance do not display any skew or shift compared to those from the ordinary CMB patches, and the distance from all 3 groups are less than 32.} 
\label{peakfromminima}
\end{center}
\end{figure}

%The Cold Spot {\it per se} does not induce peak shifts, {\it i.e.} 
One should note that a decrement, in particular, an extended decrement in CMB temperature does not cause peak shifts, i.e. there is no correlation between a decrement in CMB temperature and peaks leaning towards smaller $\l$s. To see this, we conduct CMB simulations with \planck best-fit $\Lambda$CDM model, in total we simulate 4,000 realizations of full-sky CMB maps (with 5-arcmin angular resolution). We use spherical Mexican hat wavelet (SMHW) \citep{smhw,vielva} to analyze the set of data by convolving the SMHW on the whole simulated map via 
\begin{equation}
w(R,p)=\sumlm{\alm W_\ell(R) \ylm(p)},    
\end{equation}
where $\ylm$ is the spherical harmonics, $\alm$ is the spherical harmonic coefficients of the simulated map, $p$ is the position/pixel, $W_\ell(R)$ is the SMHW window function at scale $R$, and $w(R,p)$ is the SMHW coefficient. We adopt $R=300$ arcmin and found 119 simulated cold spots that have the SMHW coefficients smaller than that of the \planck CS. We further use the CS lowest temperature as the criterion to eliminate 13, so that the 106 cold spots have deeper dip than the CS (see one example in top left panel of Fig.\ref{peakfromminima}). We place the cold spots at the center of $20\times20$ deg$^2$ patches and plot in Fig.\ref{peakfromminima} the normalized histograms of the 3 peak positions and of the distance from the \planck best-fit positions $(\ell^{(1)},\ell^{(2)},\ell^{(3)})=(220.0, 537.5, 810.8)$ \citep{planckps2}. For comparison, we plot those from 100 patches from the \planck lensing simulations. We further shift all the cold spots by 10 deg (top right panel as an example) so that roughly half of the cold spot is on the edge of the patch to mimic the CS patch and plot the histograms. One can see the normalized histograms of the peak positions and the distance of the full- and half-cold spot patches do not show any significant skew or shift compared to those from the ordinary CMB patches.

Moreover, extended cold areas with size $> 2^\circ$ are ubiquitous in CMB, which manifest themselves in the angular power spectrum at $\l < 90$, outside the range of the acoustic peaks. Thus, a more straightforward way to demonstrate the temperature dip in CS itself doesn't cause peak shifts is that we subtract the CS shape from the patch as follows: we take only the upper left quarter of the CS patch, where the CS is located, and set the temperature of the rest of the patch zero. Then it is smoothed with $\sigma=1^\circ$, with which the CS patch is subtracted. The patch now is without the prominent dip, shown in the top right panel in Figure \ref{coldspot}, and the corresponding power spectrum is shown with black dot in the middle of the same Figure. One can see the CS decrement is mitigated with power spectrum being modified only at $\ell < 180$.

%Apart from the famed Cold Spot (at $-4.7\sigma$ for SMHW $R=300$ arcmin), we list the patches containing the extrema in the \planck paper \citep{planckstat} at 4.2, 3.3, 3.3, $-3.1$, $-3.1$ $-3.2\sigma$, and we test the peak distances at  6.10, 9.39, 12.98, 2.08, 13.03, 10.13, respectively\footnote{Four of the 6 extrema are not in the 45 patches, but we still use their mean for the calculation of distance.}. One can see that all distances of the 6 extrema are below the mean from the 45 patches, 15.10, and thus one can conclude that there is no correlation between a cold area and peak shifts. 

To check the significance of the distance of the CS patch, we calculate the distance of the peaks from the 9000 patches from the 100 CMB realizations. In the bottom panel of Figure \ref{distance}, the black curve is the histogram of the SMICA map and the red 9000 simulated patches. Out of 9000 patches, only 2 with the distance larger than that of the CS patch. If one considers, however, synchronous shift towards smaller multipole numbers, there is none out of 9000 (where the highest distance with all peaks shifting towards large scales is 36.91), pushing the significance to less than $1.11\times 10^{-4}$. With the CS patch taken on the equator $\phi=[180^\circ,200^\circ]$ of the rotated SMICA map with $\theta=+67.5^\circ$, there is another slight shift from the Cold Spot patch taken at the same position from $+66^\circ$-rotated map, where the peak distance is 38.4 and shifts towards smaller multipole numbers. That an adjacent area also has the significance less than $1.11\times 10^{-4}$ shows the CS patch is not a fluke.

Below we discuss possible errors affecting the angular power spectrum. Since we compare the distance of the Cold Spot patch with those of the simulations including lensing and Doppler boosting effects, one can rule out both effect. The other common culprit is the foreground residual. One should note, however, that the patch is at high Galactic latitude, where the foreground contamination is relatively low. To demonstrate that even the foreground doesn't affect much of the peak positions, we also plot the angular power spectrum of the same field from \planck 143GHz channel map in middle panel of Figure \ref{coldspot}, which is only corrected for NPB effect. The shoot-up in power is due to the deconvolution on both the foregrounds and noise. One can see the peak positions shift very little even with the presence of the foregrounds, let alone foreground residual after the cleaning. We also plot the power spectra of the Cold Spot patch from the foreground-subtracted maps at 100, 143, 217 GHz from SEVEM method in the bottom panel, and they show the same trend. The summary of the variance of maps and estimated peak positions of the CS patch is in Table 1.

\begin{table}
\begin{tabular}{ |p{4.4cm}||p{3.3cm}| }
 \hline
 \multicolumn{2}{|c|}{Various maps} \\
 \hline
 Map & Variance from 45 patches \\
 \hline
 SMICA   & 236.79  \\
 NILC   &  233.45 \\
 SEVEM   &227.37 \\
 Commander   &232.07 \\
 143 GHz & 226.56 \\
 
 \hline
 \end{tabular}
 
 \vspace{0.3cm}

\begin{tabular}{ |p{2.5cm}||p{1.5cm}|p{1.5cm}|p{1.5cm}|  }
 \hline
 \multicolumn{4}{|c|}{CS patches from various maps} \\
 \hline
 Patch & 1st peak & 2nd peak &3rd peak\\
 \hline
 SMICA   & 206.08   & 504.76&   788.73\\
 NILC   &  204.29  & 504.78   & 791.60\\
 SEVEM   &205.84 & 505.95&  791.66\\
 Commander   &205.56 & 505.15&  791.02\\
 SMICA (no CS)&   211.67  & 502.75 &791.28 \\
 143 GHz & 206.31  & 504.27   & 792.03\\
 
 \hline
 \end{tabular}
 
 \vspace{0.3cm}
 
 \begin{tabular}{ |p{2.5cm}||p{1.5cm}|p{1.5cm}|p{1.5cm}|  }

 \hline
  \multicolumn{4}{|c|}{Foreground-subtracted CS
  Patches from SEVEM method} \\
 \hline
 patch & 1st peak & 2nd peak &3rd peak\\
 \hline
 100 GHz  & 208.40   & 508.15&   796.05\\
 143 GHz   &  205.48  & 507.09   & 790.67\\
 217 GHz   &206.20 & 505.93&  796.52\\

 \hline
 
\end{tabular}
\caption{Top Table shows the variance from the 45 patches of various maps, the middle and bottom tables show the estimated peak positions of the CS patches from various maps.}
\label{table:1}
\end{table}

%\begin{figure}
%\centering
%\epsfig{file=dl_smica.eps,width=4cm}
%\epsfig{file=dl_nilc.eps,width=4cm}
%\epsfig{file=dl_sevem.eps,width=4cm}
%\epsfig{file=dl_commander.eps,width=4cm}
%\epsfig{file=sigma_del27.eps,width=4cm}
%\epsfig{file=histo.var_fg.eps,width=4cm}
%\caption{We plot the angular power spectra for the \planck 4 maps (top 4 panels). The upper part in each panel is the mean power spectrum (big red dot) from the 45 patches (black dot) and the \planck best-fit $\Lambda$CDM power spectrum (black curve), whereas the lower part (big black dot) is the absolute difference between the mean and the \planck best-fit spectrum. The lower left panel is the dispersion of the angular power spectra from the 45 patches of the 4 Planck maps (orange SMICA, cyan NILC, red SEVEM, green C-R) and those from the 100 simulations (black dot). The lower right panel shows the difference in variance when the 100 CMB realizations are added with 3\% foregrounds.}
%\label{meanpower}
%\end{figure}

Even if there is still unknown residual after foreground cleaning, we can use analytical approach to estimate the error budget on the peak positions. We assume any unknown foreground or systematic residual has power-law power spectrum $B \ell^{-\lambda}$, then we can model the residual and an acoustic peak at $\ell_0$ with amplitude $A$ and spread $\sigma$ via $D_\ell=A \exp[-(\ell-\ell_0)^2/2\sigma^2]+ B \ell(\ell+1) \ell^{-\lambda} \exp[\ell(\ell+1)b^2]/2\pi $, where $b$ is the beam transfer function (see Appendix C). It is easy to see that, for $\lambda < 2$, the resultant $D_\ell$ at high $\l$ is dominated by $B \ell^{2-\lambda} \exp[\ell(\ell+1)b^2]/2\pi$. In the \planck SMICA, NILC and SEVEM patches, in particularly the patch containing the Cold Spot, the shoot-up is not seen at high $\l$, indicating the residual coefficient $B$ has to be extremely small to counter both the deconvolution $ \exp[\ell(\ell+1)b^2]$ and $\ell^{2-\lambda}$. Thus peak shift from a small $B$ is negligible. For $\lambda >2 $, on the other hand, the peak shift is approximately proportional to $ B\sigma^2 /A \ell_0^{\lambda-1}$. The $B$ will have to be much higher than the best-fit $\Lambda$CDM model at low $\ell$ to have any significant peak shift, which again is not seen in the angular power spectrum of the CS patch, either. For $\lambda=2$ the peak positions are mostly unchanged because the $\l(\l+1)$ cancels out the residual.

\begin{figure}[ht]
\begin{center}
\includegraphics[width=0.48\textwidth]{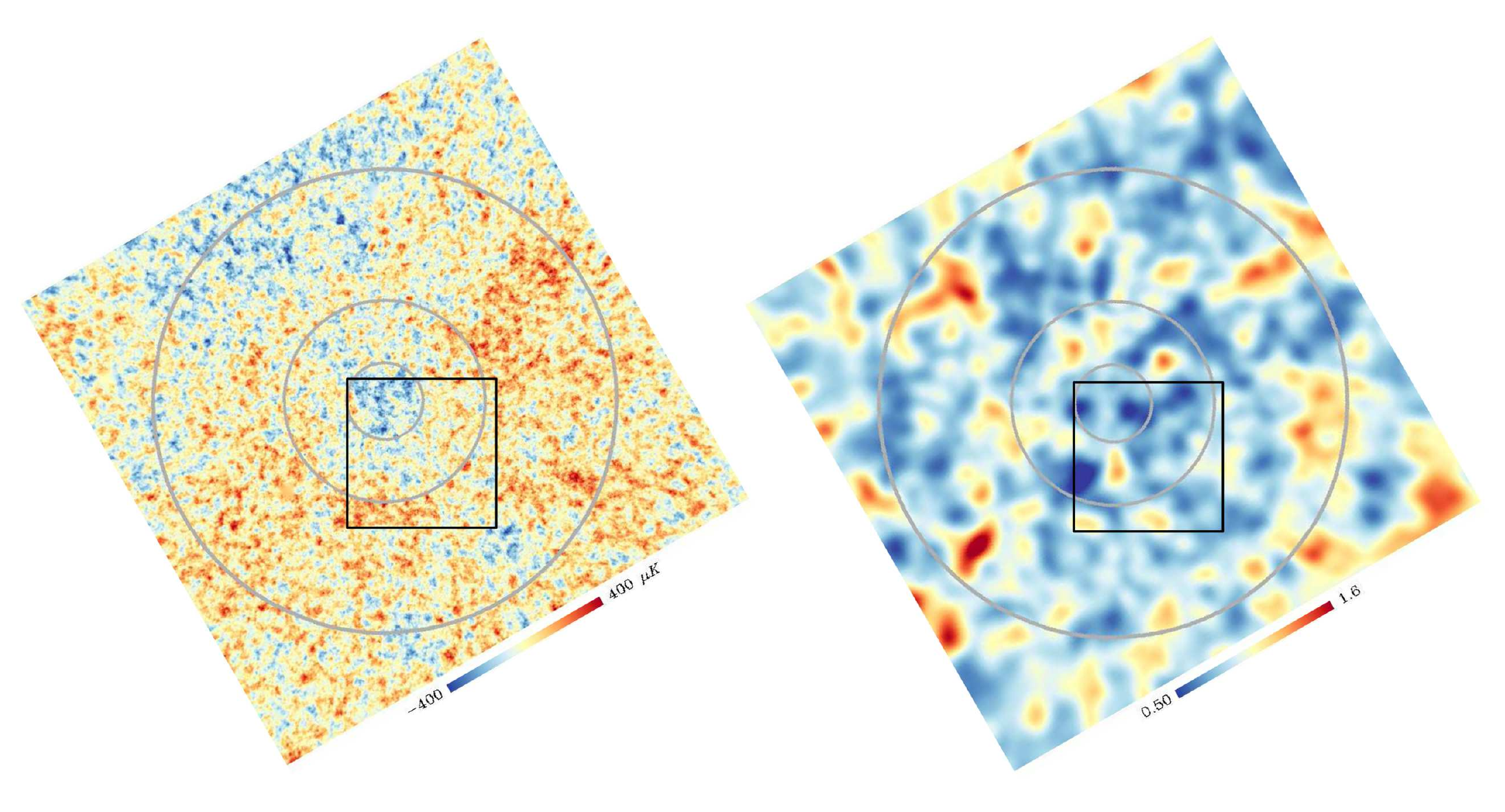}
\caption{The SMICA field centred at the CS (left) and the corresponding WISE-2MASS field (right), both panels taken from \citet{szapudiproc} (see also \citet{finelli}). The gray circles from inner to outer correspond to radii of 5$^\circ$, 14$^\circ$ and 29$^\circ$, respectively. The black square indicates our CS patch as shown in top left panel of Figure \ref{coldspot}. As is stated in \citet{szapudiproc}: ''{\it the size of the underdense region is surprisingly large: it is detected to $\sim$20$^\circ$ with high ($\geq 5\sigma$)
significance}''. One can see the surprisingly large underdense region (dark blue region) is mostly included in the square patch.}
\label{wise2mass}
\end{center}
\end{figure}

\section{Discussion} \label{sec:discussion}
Investigation of the CMB CS is well documented. Since its discovery by \citet{vielva}, there have been proposals to explain such deviation from Gaussianity: foreground \citep{cruz06,hansen}, multiple voids \citep{multivoid}, supervoid \citep{inoue06,inoue07}, cosmic texture \citep{zhao,cruz}, adiabatic perturbation on the last scattering surface \citep{valkenburg}. In particular, a supervoid is found with the size $\sim$220$h^{-1}$ Mpc to explain the lower than usual temperature by \citet{voidszapudi}. It is, however, refuted later that the model of a supervoid alone cannot explain the Cold Spot decrement \citep{nadathur,finelli,mackenzie,marcos}. Nevertheless, in trying to justify such model, \citet{voidszapudi} use WISE-2MASS catalog to identify underdensities with $z<0.3$, from which a surprisingly large underdense region $\sim$20$^\circ$ close to the CS with high ($\geq 5\sigma$) significance is detected \citep{szapudiproc}. In Figure \ref{wise2mass} we reproduce the figure 2 from \citet{szapudiproc}, and rotate 29$^\circ$ counterclockwise. The black square indicates exactly the CS patch we are investigating in this paper. One can clearly see that the large underdense region (dark blue) in the right panel is mostly included in the square.

The combination of lower than usual temperature of the CS (significance $\sim 0.01$) and larger than usual shifts towards large scales in the acoustic peaks of the patch ($1.1 \times 10^{-4}$) results in a 4.73$\sigma$ (significance $p\sim1.11\times 10^{-6}$) observation against the $\Lambda$CDM model. With another 5$\sigma$ significance large underdense region in the patch, there are 3 different anomalies in one patch that requires explanation. One should note that voids (either multiple voids or a supervoid) can't shift the acoustic peaks.

Here we propose one of the possible accounts that can explain simultaneously for these anomalies: there is some localized unknown energy in the transverse direction to stretch the space around the CS region. When space is stretched, the wavelength of sinusoidal waves are elongated so that the peaks in the harmonic domain are shifted towards larger scales, and the number density of photons is reduced, while redshifting reduces their frequency, so that the temperature is dragged down. Such stretching of space could also make the density at low redshift lower. If the unknown energy is from dark energy, this implies the existence of inhomogeneity of the dark energy\citep{deputter,blomqvist08,blomqvist10}.

%Another tantalizing evidence for such proposal is the so-called Cold Spot Repeller in \citet{csrepeller}, which finds 

%%%%%%%%%%%%%%%%%%%%%%%%%%%%%%%%%%%%%%%%%%%%%%%%%%%%%%%%%%%%%%%%%%%%%%%%%%%%%%%
\acknowledgments
This work is supported by ROC MoST grant 103-2112-M-001-028. The author acknowledges the use of ESA Planck Legacy Archive data \footnote{\tt http://pla.esac.esa.int/pla/\#home},   \healpix\footnote{\tt http://healpix.jpl.nasa.gov/} package \citep{healpix}  and \glesp\footnote{\tt http://www.glesp.nbi.dk/} package \citep{glesp}. The author would like to thank D.Spergel, P.Naselsky, A.Lewis, O.Dor\'e and S.Suyu for useful discussions and suggestions, also  Ya-Shiuan Lin for helping to produce top panel of Figure \ref{distance}.

\appendix
\section{Non-periodic boundary (NPB) correction for patches with beam convolution}
Fourier transform presumes the signal is periodic such that the two ends are continuous, thus when one deals with real data it is inevitable to encounter jump discontinuity, which induces extra power in the power spectrum. Although the data are mostly discrete, the NPB effect nevertheless persists and should be taken into account as systematic error. A patch taken from a full-sky map has such intrinsic effect, which is uncorrectable because NPB is part of the morphology. The effect, however, is further aggravated when there is beam convolution. As the statistics and comparisons in this paper are made between beam-convolved \planck data and simulations with best-fit $\Lambda$CDM model (without beam convolution), it is necessary to study the effect because the power spectrum of a beam-convolved patch has more induced power than the smoothed power spectrum of the non-periodic patch. 

To estimate the correction, we simulate 50 full-sky CMB maps and take 30 deg$^2$ patches centered at the equator (in total 600 patches). We take the central 20 deg$^2$ part directly from the 30 deg$^2$ patches, and denote their mean power spectrum as $C_\ell^A$. Then we use 5 arcmin to convolve the 30 deg$^2$ patches and choose the central 20 deg$^2$ part, the mean power spectrum of which is denoted $C_\ell^B$. The correction is therefore $C_\ell^B - b^2 C_\ell^A$, where $b$ is the beam transfer function. One finds that the extra power due to 5-arcmin beam convolution is a rather smooth curve and can be best-fitted to be $C^{\rm NPB}_\l=1.21\, \ell^{-1.38}$  ($\mu$K$^2$) from $\ell \geq 300$. And for 143GHz band map with beam transfer function 7.30 arcmin \citep{planckhfibeam} the correction for NPB is $C^{\rm NPB}_\l=34.4\, \ell^{-1.80}$  ($\mu$K$^2$).

\section{Cross-power spectrum on the \planck data}
Throughout the paper we use cross-power spectrum (XPS) on the \planck data : 4 CMB maps, 143GHz channel map, and the foreground-subtracted maps at 100, 143, and 217 GHz from the SEVEM method, to extract the peak position. The XPS is a quadratic estimator between two maps that can provide unbiased estimate of the underlying power spectrum of the correlated signals, and, at the same time, reduce the uncorrelated ones. We take from \planck Legacy Archive the FULL MISSION RINGHALF1 and FULL MISSION RINGHALF2 maps for SMICA, NILC, SEVEM, Commander and 143GHz map. Since the patches from RINGHALF1 (denoted $\alpha$) and RINGHALF2 (denoted $\beta$) contain the same CMB signal but uncorrelated noise, we can employ the XPS on their Fourier modes $\alpha_{\bi k} $ and $\beta_{\bi k}$

\begin{equation}
S_k^{\alpha\beta}= \frac{1}{2} \langle  \alpha^{*}_{\bi k} \beta_{\bi k} + \beta^{*}_{\bi k} \alpha_{\bi k} \rangle,
\end{equation}
where $*$ denotes complex conjugate and the angle brackets denote average over all $k-1/2 \le |{\bi k}| < k+1/2$ for integer k. However, as \citet{xps, direct} pointed out, lack of an ensemble and non-zero chance correlation between the so-called "uncorrelated" signal results in some residual $X_k^{\alpha\beta}$ from the XPS, which has the following relation, 
\begin{equation}
\sqrt{\langle (\xk^{\alpha\beta})^2 \rangle} \simeq \frac{\sqrt{ N^\alpha_k N^\beta_k } }{\sqrt{2\pi k}},
\end{equation}
where $N^\alpha_k$ and $N^\beta_k$ are the (uncorrelated) noise power spectrum from RINGHALF1 and RINGHALF2, respectively. The noise power spectrum can be obtain by $1/2$ of the differencing of HALFRING1 and HALFRING2 patches, assuming $N^\alpha_k$ and $N^\beta_k$ are of the same level. Thus in our data pipeline, we subtract the residual $X_k^{\alpha\beta}$  from the XPS, after the correction of NPB.

\section{Foreground residual estimate}
The foreground residual (or any unknown residual or systematics) can take the form of $ B \ell^{-\lambda}$ after foreground cleaning such that we can model the residual and an acoustic peak at $\ell_0$ with amplitude $A$ and spread $\sigma$ by $D_\ell=A \exp[-(\ell-\ell_0)^2/2\sigma^2]+ B \ell(\ell+1) \ell^{-\lambda} \exp[\ell(\ell+1)b^2]/2\pi $, where $b$ is the beam transfer function. As deconvolution of the beam transfer function varies slowly (for the multipole range in our discussion) we assume $\exp[\ell(\ell+1)b^2]\simeq w$. Letting $\beta\equiv \pi A/\sigma^2$ and $\epsilon \equiv B/\beta$, one can solve for the peak position for $\lambda=0$, 1, 2:  $\ell\simeq \ell_0 (1-\epsilon w)^{-1}$, $\ell \simeq \ell_0 + \epsilon w/ 2$, and $\ell \simeq \ell_0$, respectively. Further assuming the peak position not far from $\ell_0$ for $\lambda=3$, $\ell\sim\ell_0-\epsilon w/2\ell_0^2$.

For $\lambda=0$ the peak position is shifted by $\ell_0 \epsilon w (1-\epsilon w)^{-1}$, which could be large enough to be the source of the anomaly. One should note, however, that if there is any residual with $\lambda=0$ (flat spectrum), it should be mixed with the pixel noise and is duly subtracted in our data processing. For $\lambda=1$, the peak shift $\epsilon w/2$ is inversely proportional to $\beta$, where $\beta$ is 1.79, 0.576, 0.446 for 1st, 2nd and 3rd peak of best-fit $\Lambda$CDM model, respectively. For the shift to be significant, $B$ has to be of the same order of $\beta$. Such high level of residual $B$, together with $(\ell+1) \exp[\ell(\ell+1)b^2]$ would have caused the power to shoot up at high $\ell$, but in reality the power spectrum of the Cold Spot patch in the middle panel of Figure \ref{coldspot} shows no such shoot-up at all. For $\lambda=2$ the peak position is mostly static as the $\ell(\ell+1)$ cancels out the residual. 

The case of $\lambda=3$ ($\lambda > 2$) is different from the previous cases in that the power index of the residual part in $D_\ell$ is negative ($\sim \ell^{-1}$). The peak shift, inversely proportional to $\ell_0^2$, is less than unity even when $B$ is as high as 9000, making the residual at the same level of best-fit $\Lambda$CDM model quadrupole.

Although the foreground residual is unlikely to have power law with integer index, the analytical forms display the following trend: for $\lambda < 2$ the resultant $D_\ell$ is dominated by $B \ell^{2-\lambda} \exp[\ell(\ell+1)b^2]$ at high $\ell$. In the \planck data, in particularly the patch containing the Cold Spot, the shoot-up is not seen at high $\ell$, indicating the residual coefficient $B$ has to be quite small to counter the deconvolution and $\ell^{2-\lambda}$. The peak shift is thus negligible. For $\lambda >2 $, on the other hand, the peak shift is approximately proportional to $ \epsilon  w/\ell_0^{\lambda-1}$. The $B$ will have to be much higher than the best-fit $\Lambda$CDM model at low $\ell$ to have any significant peak shift, which again is not seen in the power spectrum in the middle of Figure \ref{coldspot}, either.

\expandafter\ifx\csname natexlab\endcsname\relax\def\natexlab#1{#1}\fi

\newcommand{\combib}[3]{\bibitem[{#1}({#2})]{#3}} %apj
%
% for authors
%
\newcommand{\autetal}[2]{{#1,\ #2., et al.,}}
\newcommand{\aut}[2]{{#1,\ #2.,}}
\newcommand{\saut}[2]{{#1,\ #2.,}}
\newcommand{\laut}[2]{{#1,\ #2.,}}
\newcommand{\coll}[1]{#1,}

%
% for papers
%
% reference for papers: 1title, 2journal, 3vol, 4page, 5year, 6astro-ph
\newcommand{\refs}[6]{#5, #2, #3, #4} %apj
\newcommand{\unrefs}[6]{#5 #2 #3 #4 (#6)}  %apj

%
% for books and proceedings
%
% reference for books: 1title, 2press, 3editor, 4edition, 5year,
% 6astro-ph

\newcommand{\book}[6]{#5, #1, #2, #3}

\def\apj{ApJ}
\def\apjl{ApJL}
\def\mn{MNRAS}
\def\nature{Nature}
\def\aa{A\&A}
\def\aas{A\&A Supplement}
\def\prl{Phys.\ Rev.\ Lett.}
\def\prd{Phys.\ Rev.\ D}
\def\pr{Phys.\ Rep.}
\def\ijmpd{Int. J. Mod. Phys. D}
\def\jcap{J. Cosmo. Astropar.}

%\combib{Arnau, Alliaga \& Saez}{2002}{arnau}\aut{Arnau}{J.V} \aut{Aliaga}{A.M} \laut{Saez}{M}\refs{Non-circular rotating beams and CMB experiments}{}{}{}{2001}{astro-ph/0112035}

\end{document}